\documentclass[journal]{IEEEtran}
\usepackage{amssymb}
\usepackage[cmex10]{amsmath}
\usepackage{stfloats}
\usepackage{graphicx}
\usepackage{subfigure}
\usepackage{tabularx}
\usepackage{epsfig,epsf,color,balance,cite}
\usepackage{verbatim}
\usepackage{url}
\usepackage{bm}
\usepackage{mathrsfs}
\usepackage{amsthm}
\DeclareMathAlphabet\mathbfcal{OMS}{cmsy}{b}{n}

\newtheorem{theorem}{Theorem}

\newtheorem{lemma}{Lemma}
\newtheorem{remark}{Remark}
\usepackage{algorithm}
\usepackage{algpseudocode}
\usepackage{graphicx}
\usepackage{epstopdf}
\IEEEoverridecommandlockouts

\begin{document}

\title{Coding-Enhanced Cooperative Jamming for\\ Secret Communication: The MIMO Case}

\author{
	\IEEEauthorblockN{Hao Xu, \emph{Member, IEEE}\IEEEauthorrefmark{0},    
		Kai-Kit Wong, \emph{Fellow, IEEE}\IEEEauthorrefmark{0},
		Yinfei Xu, \emph{Member, IEEE}\IEEEauthorrefmark{0},\\
		and
		Giuseppe Caire, \emph{Fellow, IEEE}\IEEEauthorrefmark{0}
	}
	\thanks{
		The work of H. Xu and K. K. Wong is supported by the European Union's Horizon 2020 Research and Innovation Programme under Marie Skłodowska-Curie Grant No. 101024636.
		The work of Y. Xu is supported in part by the National Natural Science Foundation of China under Grant 62371119, and in part by the Zhi Shan Young Scholar Program of Southeast University.
		The work of G. Caire is supported by the Gottfried Wilhelm Leibniz-Preis 2021 of the German Science Foundation (DFG).
		
		H. Xu and K.-K. Wong are with the Department of Electronic and Electrical Engineering, University College London, London WC1E7JE, UK (e-mail: hao.xu@ucl.ac.uk; kai-kit.wong@ucl.ac.uk).
		
		Yinfei Xu is with the School of Information Science and Engineering, Southeast University, Nanjing 210096, China (e-mail: yinfeixu@seu.edu.cn).
		
		G. Caire is with the Faculty of Electrical Engineering and Computer Science at the Technical University of Berlin, 10587 Berlin, Germany (e-mail: caire@tu-berlin.de).
		
		Corresponding author: Giuseppe Caire.
	}
}

\maketitle

\begin{abstract}
This paper considers a Gaussian multi-input multi-output (MIMO) wiretap 
channel with a legitimate transmitter, a legitimate receiver (Bob), an eavesdropper (Eve), and a cooperative jammer.
All nodes may be equipped with multiple antennas.
Traditionally, the jammer transmits Gaussian noise (GN) to enhance the security.
However, using this approach, the jamming signal interferes not only with Eve but also with Bob.
In this paper, besides the GN strategy, we assume that the jammer can also choose to use the encoded jammer (EJ) strategy, i.e., instead of GN, it transmits a codeword from an appropriate codebook. 
In certain conditions, the EJ scheme enables Bob to decode the jamming codeword and thus cancel the interference, while Eve remains unable to do so even if it knows all the codebooks.
We first derive an inner bound on the system's secrecy rate under the strong secrecy metric, 
and then consider the maximization this bound through precoder design in a computationally efficient manner.
In the single-input multi-output (SIMO) case, we prove that although non-convex, the power control problems can be optimally solved for both GN and EJ schemes. In the MIMO case, we propose to solve the problems using the matrix simultaneous diagonalization (SD) technique, which requires quite a low computational complexity.
Simulation results show that by introducing a cooperative jammer with coding capability, and allowing it to switch between the GN and EJ schemes, a dramatic increase in the secrecy rate can be achieved.
In addition, the proposed algorithms can significantly outperform the 
current state of the art benchmarks in terms of both secrecy rate and computation time.
\end{abstract}

\IEEEpeerreviewmaketitle

\section{Introduction}
\label{section1}

Nowadays, more and more private information is transmitted over the air, leading to a growing need for enhanced security measures in 5G and forthcoming 6G mobile systems \cite{Ericsson2022}.
In addressing this concern, information theoretic security, often referred to as physical layer security, has emerged as a promising approach to ensure secure communication \cite{ wu2018survey}.
As shown in \cite{csiszar1978broadcast}, in a wiretap channel, positive secrecy rate can be obtained only when the legitimate receiver (Bob for short) experiences better channel condition than the eavesdropper (Eve for short).
Otherwise, it is impossible for the transmitter to communicate securely with its intended receiver.
Consequently, numerous studies have been dedicated to enhancing the channel conditions of the legitimate link through various techniques, among which jamming has been proven to be a powerful method \cite{atallah2015survey, huo2017jamming}, and has found widespread application in various communication systems \cite{lee2018uav, sun2022ris, chen2022security}.

In the information theoretic literature, the most commonly considered jamming scheme consists of transmitting Gaussian noise (GN) \cite{goel2008guaranteeing, tekin2008general, fakoorian2011solutions}.
However, in this scheme, the jamming signal interferes not only with Eve, but also with Bob.
Although in some cases it is possible to apply signal processing techniques to completely eliminate the interference caused to Bob \cite{wolf2010zero, hu2018cooperative, li2014secrecy, chu2014secrecy}, there are limitations to these methods.
For example, in \cite{wolf2010zero} and \cite{hu2018cooperative}, zero-forcing beamforming was applied to avoid interference to Bob and, in \cite{li2014secrecy} and \cite{chu2014secrecy}, the covariance matrix of the jammer was designed to ensure that its signal lies in the null-space of the channel from itself to Bob.
However, these methods are not generally applicable to the multi-input multi-output (MIMO) case, because to maintain a non-empty null-space for the channel from the jammer to Bob, the jammer must have more antennas than Bob (see \cite{wolf2010zero, hu2018cooperative, li2014secrecy, chu2014secrecy}).

To avoid interfering with Bob and ensure compatibility with the general MIMO scenarios, an encoded jammer (EJ) transmitting codewords from an appropriately designed codebook rather than GN, 
can be used.  
Under certain conditions, the EJ scheme enables Bob to decode the jamming codeword 
and thus cancel the interference, while Eve remains unable to do so even if it possesses all codebooks.
It has been proven by \cite{tang2008gaussian, tang2011interference, he2014providing} that, using the EJ scheme, it is possible to improve the secrecy rate over the GN scheme.
Specifically, in \cite{tang2008gaussian} and \cite{tang2011interference}, the achievable secrecy rate of a discrete memoryless (DM) wiretap channel with an encoded jammer was analyzed, and the secrecy performance was also verified in a single-antenna Gaussian wiretap channel by using the Gaussian random coding strategy.
In \cite{he2014providing}, a similar scalar Gaussian wiretap channel was considered, and it was shown that by using lattice structured codes, the achievable secrecy rate does not saturate at high signal-to-noise ratio (SNR).

However, it should be noticed that the secrecy performance of the EJ scheme may not always surpass that of the GN method, as it depends on the specific channel conditions.
This arises from the fact that, using the EJ scheme, both the legitimate transmitter and the jammer transmit encoded messages. 
To ensure successful decoding of all the information by Bob while preventing Eve from decoding, additional constraints will be imposed on the secrecy rate.
In Remark~\ref{remark1} of this paper, we delve into a specific channel realization and provide a more detailed analysis in this regard. 
Consequently, it is crucial to thoroughly investigate and compare the performance of the GN and EJ schemes across different channel conditions and system configurations. 
Notably, this particular problem remains unexplored in the context of multi-antenna systems, which serves as the primary motivation behind this work.

This paper considers a Gaussian MIMO wiretap channel with a legitimate transmitter, a cooperative jammer, a legitimate receiver, and an eavesdropper.
The jammer is assumed to have the ability to switch between the GN and EJ schemes, depending on which one provides better jamming performance.
This system has rarely been studied, and its achievable region under the strong secrecy metric is still unknown.
Therefore, we aim to investigate its secrecy performance, in particular the gain obtained by introducing the EJ scheme.
Also, we propose novel precoder design algorithms that achieve better performance with lower computational complexity with respect to the current state of the art approaches.
The main contributions are summarized below.

$\bullet$ We first derive an inner bound on the system secrecy rate under the strong secrecy metric.\footnote{The {\em weak secrecy} criterion is characterized by the {\em {information leakage rate}}. However, a vanishing information leakage rate does not imply that a vanishing number of information bits of the secret message are leaked, because the length of the message in bits grows linearly with the block length. To address this issue, {\em strong secrecy} was introduced in \cite{maurer2000information, csiszar2000common}, by considering directly the {\em information leakage} without normalization by the block length, and has been considered a safer secrecy metric  \cite{bloch2013strong}.}
Note that the achievable secrecy rate of a DM case in a similar context was analyzed under weak secrecy in \cite{tang2011interference} by discussing different interference levels.
To gain deeper insights into various jamming schemes and facilitate our analysis, we take a different approach from \cite{tang2011interference} by separately investigating the achievable secrecy rates of the GN and EJ strategies.
Interestingly, by examining a real and scalar Gaussian scenario (see Subsection~\ref{special_case}), we show that although considering the strong secrecy metric, the derived inner bound aligns with that given in \cite{tang2011interference}.

$\bullet$ To enhance the security, we maximize the derived inner bound by precoder design.
This problem can be decomposed into two subproblems, each aiming to maximize the secrecy rate when the GN and EJ schemes are employed.
We start from the single-input multi-output (SIMO) case, where both the legitimate transmitter and the jammer have one antenna, while Bob and Eve have multiple antennas.
Although non-convex, we prove that the power control problems for different jamming schemes can be optimally solved, and the globally optimal solutions can be obtained essentially in closed form.

$\bullet$ For the general MIMO case, both subproblems become much more complicated and closed-form solutions do not appear to be possible.
A meaningful approach to precoder design consists of finding good ``heuristic solutions'', i.e., feasible points that yield good results in terms of the objective function. 
In the following, we shall refer to ``solution'' in this sense.	
As we will show in Section~\ref{SDLC_MIMO} and the simulation, a heuristic solution to each subproblem can be achieved by first obtaining its convex approximation using the majorization minimization (MM) technique, and then iteratively optimizing this approximation using some standard tools such as the interior-point method.
However, this approach requires the solution of log-determinant optimizations in each iteration, yielding extremely high complexity. 
To address the issue, we propose a novel approach based on matrix simultaneous diagonalization (SD). 
Using this technique, we show that the precoder design problems associated with different jamming schemes can be efficiently solved with comparable performance (often better) and 
much less computational complexity than the MM-based method.
Simulation results show that by allowing the jammer to switch between the GN and EJ schemes to leverage the strengths of both approaches, a remarkable increase in the secrecy rate can be achieved.
Moreover, the proposed algorithms can significantly outperform the benchmarks in terms of both secrecy rate and computation time.

The rest of this paper is organized as follows. 
In Section~\ref{GV_MAC-WT}, the system model is provided and the problem is formulated.
In Section~\ref{power_control_SIMO} and Section~\ref{SDLC_MIMO}, we respectively consider the power control and precoder design problems for the SIMO and MIMO cases.
Simulation results are presented in Section~\ref{simulation} before conclusions in Section~\ref{conclusion}.

Notations: $\mathbb R$ and $\mathbb C$ respectively represent the real and complex spaces. 
Boldface lower and upper case letters are used to denote vectors and matrices.
${\bm I}_N$ stands for the $N \times N$ dimensional identity matrix and $\bm 0$ denotes the all-zero vector or matrix.
Superscript $(\cdot)^H$ means conjugate transpose and $[\cdot]^+ \triangleq \max (\cdot,0)$.
The logarithm function $\log$ is base $2$.

Before going into the main part, we would like to present some equations of matrices, since they will be used frequently in this paper. For any matrices $\bm O_1$ and $\bm O_2$, if the dimensions match, we have \cite{petersen2008matrix}
\begin{align}
	| \bm O_1 \bm O_2 + \bm I | & = | \bm O_2 \bm O_1 + \bm I |, \label{O1O2_1}\\
	|\bm O_1 \bm O_2| & = |\bm O_1| |\bm O_2|, \label{O1O2_2} \\
	{\text {tr}} (\bm O_1 \bm O_2) & = {\text {tr}} (\bm O_2 \bm O_1). \label{O1O2_3}
\end{align}


\section{System Model and Problem Formulation}
\label{GV_MAC-WT}

\subsection{System Model}
We consider a Gaussian MIMO wiretap channel with a legitimate transmitter (Tx1), a legitimate receiver (Bob), an eavesdropper (Eve), and an additional transmitter (Tx2) that serves as a cooperative jammer to improve the secrecy performance of Tx1.
Transmitter $k \in \{ 1, 2 \}$, Bob, and Eve are respectively equipped with $N_k$, $N_{\text b}$, and $N_{\text e}$ antennas.
Let ${\bm x}_k \in {\mathbb C}^{N_k \times 1}$ denote the signal vector of transmitter $k$ and we assume Gaussian channel input, i.e., $\bm x_k \sim {\cal CN}(\bm 0, \bm Q_k)$.
Here $\bm Q_k$ is the covariance matrix of the signal vector and satisfies ${\text {tr}}(\bm Q_k) \leq P_k$, where $P_k$ is the maximum transmit power.
The received signals at Bob and Eve are respectively given by
\begin{align}\label{GV_YZ}
	& \bm y = \bm H_1 \bm x_1 + \bm H_2 \bm x_2 + \bm \varepsilon_{\text b},\nonumber\\
	& \bm z = \bm G_1 \bm x_1 + \bm G_2 \bm x_2 + \bm \varepsilon_{\text e},
\end{align}
where $\bm H_k \in {\mathbb C}^{N_{\text b} \times N_k}$ and $\bm G_k \in {\mathbb C}^{N_{\text e} \times N_k}$ are constant channel matrices from transmitter $k$ to Bob and Eve, and $\bm \varepsilon_{\text b} \in {\mathbb C}^{N_{\text b} \times 1}$ and $\bm \varepsilon_{\text e} \in {\mathbb C}^{N_{\text e} \times 1}$ are Gaussian noise vectors at Bob and Eve with $\bm \varepsilon_{\text b} \sim {\cal CN}(\bm 0, \bm I_{N_{\text b}})$ and $\bm \varepsilon_{\text e} \sim {\cal CN}(\bm 0, \bm I_{N_{\text e}})$.

\subsection{Problem Formulation}

In order to obtain better secrecy performance, we assume that Tx2 has the flexibility to switch between the GN and EJ modes, depending on which one provides a better jamming performance.
If the GN mode is chosen, Tx2 simply transmits Gaussian noise and its signal interferes with both Bob and Eve.
In this case, for given $\bm Q_1$ and $\bm Q_2$, the maximum secrecy rate of Tx1 under the strong secrecy metric is \cite{barros2008strong}
\begin{align}\label{SR_GN_jam}
	R_{\text {GN}} & = \left[ \log | \bm H_1 \bm Q_1 \bm H_1^H \left( \bm H_2 \bm Q_2 \bm H_2^H + \bm I_{N_{\text b}} \right)^{-1} + \bm I_{N_{\text b}} | \right. \nonumber\\
	& \left. - \log | \bm G_1 \bm Q_1 \bm G_1^H \left( \bm G_2 \bm Q_2 \bm G_2^H + \bm I_{N_{\text e}} \right)^{-1} + \bm I_{N_{\text e}} | \right]^+.
\end{align}

If the EJ mode is chosen, Tx2 jams by transmitting encoded codewords instead of Gaussian noise.
By adopting appropriate coding approaches, Bob can successfully decode the signal of Tx2 and then eliminate the interference, while Eve cannot, even if it knows the codebooks.
We show in Appendix~\ref{prove_theorem_inner} that under the EJ scheme, model (\ref{GV_YZ}) can be seen as a special case of the two-user wiretap channel, where both the users transmit secret messages.
The achievable region of such a system has been widely studied, but the capacity region remains as an open problem \cite{yassaee2010multiple, xu2023achievable}.
Therefore, in the following theorem, we provide an inner bound on the secrecy rate of Tx1 when Tx2 works on the EJ mode.
\begin{theorem}\label{theorem_inner}
	For given $\bm Q_1$ and $\bm Q_2$, if Tx2 adopts the EJ strategy for cooperative jamming, then, secrecy rate satisfying 
	\begin{equation}\label{achie_region}
		R \leq R_{\text {EJ}} \triangleq \max \{ \min \{ {\hat R}, {\tilde R} \}, {\bar R} \},
	\end{equation}
	is achievable under the strong secrecy metric, where
	\begin{align}\label{SR_hat_tilde_bar_MIMO}
		{\hat R} & = \left[ \log | \bm H_1 \bm Q_1 \bm H_1^H + \bm I_{N_{\text b}} | \right.\nonumber\\
		& \left. - \log | \bm G_1 \bm Q_1 \bm G_1^H \left( \bm G_2 \bm Q_2 \bm G_2^H + \bm I_{N_{\text e}} \right)^{-1} + \bm I_{N_{\text e}} | \right]^+,\nonumber\\
		{\tilde R} & = \left[ \log | \bm H_1 \bm Q_1 \bm H_1^H + \bm H_2 \bm Q_2 \bm H_2^H + \bm I_{N_{\text b}} | \right.\nonumber\\
		& \left. - \log | \bm G_1 \bm Q_1 \bm G_1^H + \bm G_2 \bm Q_2 \bm G_2^H + \bm I_{N_{\text e}} | \right]^+, \nonumber\\
		{\bar R} & = \left[ \log | \bm H_1 \bm Q_1 \bm H_1^H + \bm I_{N_{\text b}} | \!-\! \log | \bm G_1 \bm Q_1 \bm G_1^H + \bm I_{N_{\text e}} | \right]^+\!\!.\!\!
	\end{align}
\end{theorem}
\itshape \textbf{Proof:} \upshape
See Appendix~\ref{prove_theorem_inner}.
\hfill $\Box$

\begin{remark}\label{remark1}
	Comparing (\ref{SR_GN_jam}) with (\ref{SR_hat_tilde_bar_MIMO}), 
	we see that $R_{\text {GN}}$ and $R_{\text {EJ}}$ do not obey a fixed ordering relationship that holds in all channel conditions. 
	This is due to the fact that the two jamming schemes operate on distinct principles.
	In the GN scheme, Tx2 directly transmits an uncoded random signal, which interferes with both Bob and Eve.
	Differently, in the EJ scheme, Tx2 transmits a randomly chosen codeword from a codebook of given rate.
	By designing the coding scheme, Bob can successfully decode and eliminate the interference from Tx2, but Eve cannot even if it knows the codebooks.
	This makes it possible for the EJ scheme to outperform the GN method. 
	However, this may not always hold because to ensure that Bob can decode both the desired information and the jammer message, more constraints are imposed on the secrecy rate of Tx1 (see (\ref{SR_hat_tilde_bar_MIMO})).\hfill $\lozenge$
\end{remark}

\begin{figure}
	\centering
	\includegraphics[scale=0.47]{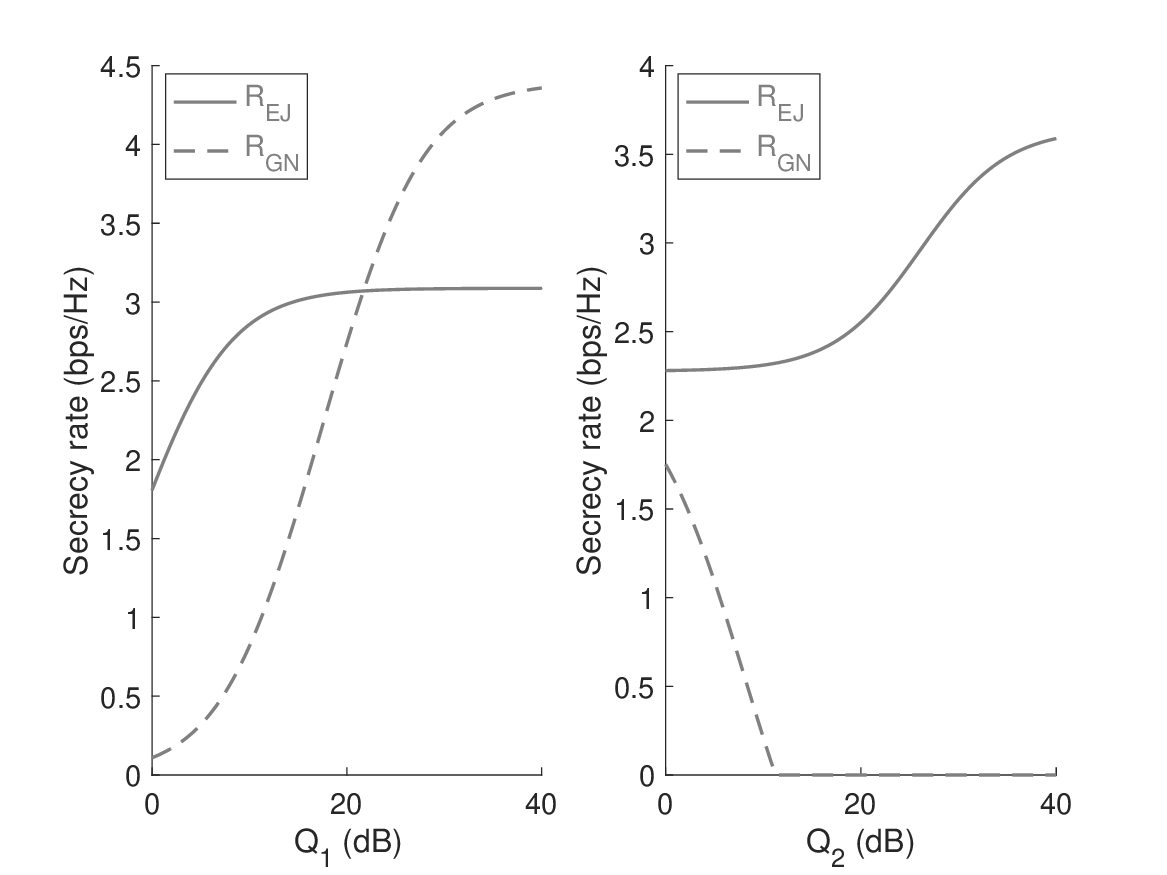}
	\caption{Left: secrecy rate versus $\bm Q_1$ with $\bm Q_2 = 20$ dB, $\bm H_1 = -0.5 + 2i$, $\bm H_2 = 0.5 - 0.5i$, $\bm G_1 = -0.5 + 0.5i$, and $\bm G_2 = - 1 - 0.5i$. Right: secrecy rate versus $\bm Q_2$ with $\bm Q_1 = 20$ dB, $\bm H_1 = -0.5 - i$, $\bm H_2 = 0.5 + 0.5i$, $\bm G_1 = 0.5i$, and $\bm G_2 = 0.2i$.}
	\label{Fig1}
\end{figure}

To get a more intuitive understanding of Remark~\ref{remark1}, we consider a system with single-antenna nodes, i.e., $N_1 = N_2 = N_{\text b} = N_{\text e} =1$, and depict $R_{\text {GN}}$ and $R_{\text {EJ}}$ for given channel realizations in Fig.~\ref{Fig1}.\footnote{Note that in the single-antenna case, the matrices $\bm Q_1$ and $\bm Q_2$ are real scalars. Therefore, in Fig.~\ref{Fig1}, $\bm Q_1 = 20$ dB means that $\bm Q_1 = 100$.}
It can be seen that under certain channel conditions, the EJ scheme exhibits superior performance over the GN scheme, but there are also scenarios where its performance is inferior to that of the GN scheme.
Driven by this observation, in this paper we consider a cooperative jammer that can choose between the GN and EJ schemes, depending on the channel state.

Based on (\ref{SR_GN_jam}) and Theorem~\ref{theorem_inner}, an inner bound (achievable lower bound) to the secrecy capacity of the Gaussian MIMO wiretap channel defined in (\ref{GV_YZ}) under the strong secrecy metric is given by 
\begin{equation}\label{C}
	{\underline C} = \max\{R_{\text {GN}}, R_{\text {EJ}}\}.
\end{equation}
In this information theoretic setting, as usual in physical layer security works, we assume that full channel state information (CSI) is known and simply determine the best strategy based on the best achievable secrecy rate, i.e., Tx2 works on the GN mode if $R_{\text {GN}} \geq R_{\text {EJ}}$, and on the EJ mode if $R_{\text {GN}} < R_{\text {EJ}}$.
Note that, as explained in Remark~\ref{remark1} and shown in Fig.~\ref{Fig1}, $R_{\text {GN}}$ and $R_{\text {EJ}}$ do not obey a fixed ordering relationship that holds in all channel states. 
Therefore, if the channel state changes and the relationship between $R_{\text {GN}}$ and $R_{\text {EJ}}$ is reversed, Tx2 can shift its working mode from one to another.
In a practical scenario, of course, lack of CSI may make things more complicated. 
However, wiretap channels in the presence of non-perfect or missing CSI are still a wide open problem in information theory and they are well beyond the scope of this paper. 
In this paper we stick to the classical setting of full CSI, based on which Tx1 and Tx2 can agree on the best strategy and use it to achieve the best secrecy rate available.

This paper aims to maximize ${\underline C}$ by optimizing the covariance matrices $\bm Q_1$ and $\bm Q_2$, and investigate the secrecy performance of Tx1 under different jamming schemes.
Obviously, the problem can be solved by separately maximizing $R_{\text {GN}}$ and $R_{\text {EJ}}$, i.e., considering the following two subproblems
\begin{align}
	\mathop {\max }\limits_{ \bm Q_1, \bm Q_2 } \quad & R_{\text {GN}} \nonumber\\
	\text{s.t.} \quad\; & \bm Q_k \succeq \bm 0,~ {\text {tr}}(\bm Q_k) \leq P_k, ~\forall~ k \in \{1, 2\}, \label{max_SR_GN}
	\\
	\mathop {\max }\limits_{ \bm Q_1, \bm Q_2 } \quad & R_{\text {EJ}}  \nonumber\\
	\text{s.t.} \quad\; & \bm Q_k \succeq \bm 0,~ {\text {tr}}(\bm Q_k) \leq P_k, ~\forall~ k \in \{1, 2\}. \label{max_SR_EJ}
\end{align}

\subsection{Special Case: Real and Scalar Gaussian Channel}
\label{special_case}
Before addressing the general problem, 
we first consider the real and scalar Gaussian channel studied in \cite[(19)]{tang2011interference}, 
which is a special case of (\ref{GV_YZ}).
We show that in this case, ${\underline C}$ in (\ref{C}) can be easily computed, and the resulting value of ${\underline C}$ is consistent with that provided in \cite[Theorem~$3$]{tang2011interference} although \cite{tang2011interference} considered the weak secrecy metric.
With appropriate notation adjustments, the channel model in \cite[(19)]{tang2011interference} can be expressed as follows
\begin{align}\label{scalar_yz}
	& y = x_1 + \sqrt{\mu} x_2 + \varepsilon_{\text b},\nonumber\\
	& z = \sqrt{\nu} x_1 + x_2 + \varepsilon_{\text e},
\end{align}
where $x_k \sim {\cal N} (0, q_k)$, $\varepsilon_{\text b}, \varepsilon_{\text e} \sim {\cal N} (0, 1)$, and $\mu$ and $\nu$ are positive real numbers.
Obviously, (\ref{scalar_yz}) is a special case of (\ref{GV_YZ}).
In this scenario, the rates in (\ref{SR_GN_jam}) and (\ref{SR_hat_tilde_bar_MIMO}) can be simplified as
\begin{align}\label{rate_scalar}
	R_{\text {GN}} & = \left[ \gamma \left( \frac{q_1}{1 + \mu q_2} \right) - \gamma \left( \frac{\nu q_1}{1 + q_2} \right) \right]^+,\nonumber\\
	{\hat R} & = \left[ \gamma (q_1) - \gamma \left( \frac{\nu q_1}{1 + q_2} \right) \right]^+,\nonumber\\
	{\tilde R} & = \left[ \gamma (q_1 + \mu q_2) - \gamma ( \nu q_1 + q_2 ) \right]^+,\nonumber\\
	{\bar R} & = \left[ \gamma (q_1) - \gamma (\nu q_1 ) \right]^+,
\end{align}
where $\gamma (\cdot) = \frac{1}{2} \log(1 + \cdot)$.
Then, ${\underline C}$ in (\ref{C}) can be rewritten as
\begin{align}\label{C_scalar}
	{\underline C} & = \max\{R_{\text {GN}}, \max \{ \min \{ {\hat R}, {\tilde R} \}, {\bar R} \}\} \nonumber\\
	& = \max\{ \max \{ R_{\text {GN}}, \min \{ {\hat R}, {\tilde R} \} \}, {\bar R} \}.
\end{align}
By checking the value of $\mu$, the expression of ${\underline C}$ can be simplified to a more concise form as shown in the following lemma. The proof is straightforward and it is omitted because of space limitation.
\begin{lemma}\label{lemma_value_R_GN}
	For the real and scalar Gaussian wiretap channel in (\ref{scalar_yz}), with fixed transmit power $q_1$ and $q_2$, the secrecy rate ${\underline C}$ in (\ref{C}) is given by (\ref{C_scalar}),	where $\max \{ R_{\text {GN}}, \min \{ {\hat R}, {\tilde R} \} \}$ can be computed as follows
	\begin{equation}\label{optimal_rate}
		\max \{ R_{\text {GN}}, \min \{ {\hat R}, {\tilde R} \} \} \!=\! \left\{\!\!\!\!\!\!\!\!
		\begin{array}{ll}
			& {\hat R}, ~~~~{\text {if}}~ \mu \geq 1 + q_1, \\
			& {\tilde R}, ~~~~{\text {if}}~ 1 \leq \mu < 1 + q_1, \\
			& R_{\text {GN}}, ~{\text {if}}~ \mu < 1.
		\end{array} \right.\!\!\!
	\end{equation}
\hfill $\Box$
\end{lemma}
It can be readily checked that Lemma~\ref{lemma_value_R_GN} is consistent with \cite[Theorem~$3$]{tang2011interference}.
In this special case, the problem of maximizing ${\underline C}$ can be easily and optimally solved (see \cite[Lemma~$1$]{tang2011interference}).
However, for the general MIMO case considered in this paper, the problem is significantly more complicated and has not 
been addressed in the existing literature. In the following sections, we focus on solving (\ref{max_SR_GN}) and (\ref{max_SR_EJ}) by separately considering the SIMO and MIMO cases.

\section{Optimal Power Control for the SIMO Case}
\label{power_control_SIMO}

In this section, we consider the SIMO case, where both the transmitters have one antenna, while Bob and Eve have multiple antennas.
In this case, the covariance matrix $\bm Q_k$ and channel matrices $\bm H_k$ and $\bm G_k$ respectively reduce to $q_k \in {\mathbb R}$, $\bm h_k \in {\mathbb C}_{N_{\text b} \times 1}$, and $\bm g_k \in {\mathbb C}_{N_{\text e} \times 1}$.
We prove that although the power control problems are non-convex, the optimal solution in closed form can be obtained for different jamming schemes.

\subsection{GN Scheme}
\label{conven_jam}
In the SIMO case, $R_{\text {GN}}$ in (\ref{SR_GN_jam}) can be rewritten as
\begin{align}\label{SR_conven_jam}
R_{\text {GN}} = & \left[ \log | q_1 \bm h_1 \bm h_1^H \left( q_2 \bm h_2 \bm h_2^H + \bm I_{N_{\text b}} \right)^{-1} + \bm I_{N_{\text b}} | \right. \nonumber\\
& \left. - \log | q_1 \bm g_1 \bm g_1^H \left( q_2 \bm g_2 \bm g_2^H + \bm I_{N_{\text e}} \right)^{-1} + \bm I_{N_{\text e}} | \right]^+.
\end{align}
If Tx2 adopts the GN jamming strategy, we consider problem (\ref{max_SR_GN}), which becomes
\begin{align}\label{max_SR_conven_jam}
	\mathop {\max }\limits_{q_1, q_2} \quad & R_{\text {GN}} \nonumber\\
	\text{s.t.} \quad\; & 0 \leq q_k \leq P_k, ~\forall~ k \in \{ 1, 2 \}. 
\end{align}
From (\ref{SR_conven_jam}), we see that problem (\ref{max_SR_conven_jam}) is non-convex.
However, we show in the following theorem that it can be optimally solved.
\begin{theorem}\label{theorem_opt_solu_conven}
	The optimal solution of problem (\ref{max_SR_conven_jam}) is $q_1^* = P_1$ and 
	\begin{equation}\label{optimal_p2_conven_jam}
	q_2^* \!=\! \left\{\!\!\!\!\!\!\!\!\!\!
	\begin{array}{ll}
	& - \frac{c}{b}, ~ {\text {if}}~ a = 0,~ b < 0,~ {\text {and}}~ 0 < - \frac{c}{b} < P_2, \\
	& \arg \!\mathop {\max }\limits_{q_2 \in \{\! P_0, P_2 \!\}}\! R_{\text {GN}}, ~{\text {if}}~ a \!>\! 0,~\!\! b^2 \!\!-\!\! 4ac \!>\! 0,~\!\! {\text {and}}~ 0 \!\!<\!\! P_0 \!\!<\!\! P_2, \\
	& \arg \!\mathop {\max }\limits_{q_2 \in \{\! 0, P_0 \!\}}\! R_{\text {GN}}, ~{\text {if}}~ a \!<\! 0,~\!\! b^2 \!-\! 4ac \!>\! 0,~\!\! {\text {and}}~ 0 \!\!<\!\! P_0 \!\!<\!\! P_2, \\
	& \arg \!\mathop {\max }\limits_{q_2 \in \{\! 0, P_2 \!\}}\! R_{\text {GN}}, ~{\text {otherwise}}, \\
	\end{array} \right.
	\end{equation}
	where $a$, $b$, and $c$ are defined in (\ref{abc}), and $P_0$ is given in (\ref{zero_point}).
\end{theorem}
\itshape \textbf{Proof:} \upshape
See Appendix \ref{prove_theorem_opt_solu_conven}.
\hfill $\Box$

Theorem~\ref{theorem_opt_solu_conven} shows that in the optimal case, Tx1 always transmits at the maximum power, while the power of Tx2 can be found by simply checking the channel condition and verifying a few possible solutions.

\subsection{EJ Scheme}
\label{IT_CJ}
In the SIMO case, ${\hat R}$, ${\tilde R}$, and ${\bar R}$ in (\ref{SR_hat_tilde_bar_MIMO}) can be rewritten as
\begin{align}\label{SR_hat_tilde_bar_SIMO}
{\hat R} & = \left[ \log | q_1 \bm h_1 \bm h_1^H + \bm I_{N_{\text b}} | \right. \nonumber\\
& \left. - \log | q_1 \bm g_1 \bm g_1^H \left( q_2 \bm g_2 \bm g_2^H + \bm I_{N_{\text e}} \right)^{-1} + \bm I_{N_{\text e}} | \right]^+, \nonumber\\
{\tilde R} & = \left[ \log | q_1 \bm h_1 \bm h_1^H + q_2 \bm h_2 \bm h_2^H + \bm I_{N_{\text b}} | \right. \nonumber\\
& \left. - \log | q_1 \bm g_1 \bm g_1^H + q_2 \bm g_2 \bm g_2^H + \bm I_{N_{\text e}} | \right]^+,\nonumber\\
{\bar R} & = \left[ \log | q_1 \bm h_1 \bm h_1^H + \bm I_{N_{\text b}} | - \log | q_1 \bm g_1 \bm g_1^H + \bm I_{N_{\text e}} | \right]^+.
\end{align}
If Tx2 operates in the EJ mode, we consider problem (\ref{max_SR_EJ}), which becomes
\begin{align}
	\mathop {\max }\limits_{q_1, q_2} \quad & \max \{ \min \{ {\hat R}, {\tilde R} \}, {\bar R} \} \nonumber\\
	\text{s.t.} \quad\; & 0 \leq q_k \leq P_k, ~\forall~ k \in \{ 1, 2 \}. \label{max_SR_new_jam}
\end{align}
Obviously, problem (\ref{max_SR_new_jam}) can be solved by separately maximizing $\min \{ {\hat R}, {\tilde R} \}$ and ${\bar R}$.
Note that (\ref{SR_hat_tilde_bar_SIMO}) implies that ${\bar R}$ can be obtained directly from either ${\hat R}$ or ${\tilde R}$ by simply letting $q_2 = 0$.
Therefore, (\ref{max_SR_new_jam}) can be equivalently simplified to
\begin{align}
	\mathop {\max }\limits_{q_1, q_2} \quad & \min \{ {\hat R}, {\tilde R} \} \nonumber\\
	\text{s.t.} \quad\; & 0 \leq q_k \leq P_k, ~\forall~ k \in \{ 1, 2 \}, \label{max_SR_new_jam2}
\end{align}
which is a max-min problem and is usually difficult to solve.
However, we prove that (\ref{max_SR_new_jam2}) can be optimally solved.
Before giving the result, we first consider the following two problems
\begin{align}
	\mathop {\max }\limits_{q_1, q_2} \quad & {\hat R} \nonumber\\
	\text{s.t.} \quad\; & {\hat P}_{1, {\text {lb}}} \leq q_1 \leq {\hat P}_{1, {\text {ub}}},~ 0 \leq q_2 \leq P_2, \label{sub_problem1}\\
	\mathop {\max }\limits_{q_1, q_2} \quad & {\tilde R} \nonumber\\
	\text{s.t.} \quad\; & {\tilde P}_{1, {\text {lb}}} \leq q_1 \leq {\tilde P}_{1, {\text {ub}}},~ 0 \leq q_2 \leq P_2, \label{sub_problem2}
\end{align}
which respectively maximize ${\hat R}$ and ${\tilde R}$.
Note that different from (\ref{max_SR_new_jam2}), new lower and upper bounds ${\hat P}_{1, {\text {lb}}}, {\hat P}_{1, {\text {ub}}}, {\tilde P}_{1, {\text {lb}}}, {\tilde P}_{1, {\text {ub}}} \in [0, P_1]$ are imposed on $q_1$ in (\ref{sub_problem1}) and (\ref{sub_problem2}), and apparently, these bounds satisfy ${\hat P}_{1, {\text {lb}}} < {\hat P}_{1, {\text {ub}}}$ and ${\tilde P}_{1, {\text {lb}}} < {\tilde P}_{1, {\text {ub}}}$.
In the following theorem we show that both (\ref{sub_problem1}) and (\ref{sub_problem2}) can be optimally solved, based on which (\ref{max_SR_new_jam2}) can then be solved.
\begin{theorem}\label{theorem_opt_p1p2_new}
	The optimal solution of problem (\ref{sub_problem1}) is 
	\begin{equation}\label{opt_p1_new_jam}
	{\hat q}_1^* = \left\{\!\!\!\!\!\!\!\!
	\begin{array}{ll}
	& {\hat P}_{1, {\text {ub}}}, ~{\text {if}}~ \bm h_1^H \bm h_1 > \bm g_1^H \left( P_2 \bm g_2 \bm g_2^H + \bm I_{N_{\text e}} \right)^{-1} \bm g_1, \\
	& {\hat P}_{1, {\text {lb}}}, ~{\text {otherwise}}.
	\end{array} \right.
	\end{equation}
	and ${\hat q}_2^* = P_2$.
	The optimal solution of problem (\ref{sub_problem2}) is 
	\begin{equation}\label{opt_p2_new_jam}
	({\tilde q}_1^*, {\tilde q}_2^*) = \arg \mathop {\max }\limits_{ (q_1, q_2) \in \left\{ ({\tilde P}_{1, {\text {lb}}}, 0), ({\tilde P}_{1, {\text {lb}}}, P_2), ({\tilde P}_{1, {\text {ub}}}, 0), ({\tilde P}_{1, {\text {ub}}}, P_2) \right\} } {\tilde R},
	\end{equation}
	which can be easily found by checking four possible solutions.
\end{theorem}
\itshape \textbf{Proof:} \upshape
See Appendix \ref{prove_theorem_opt_p1p2_new}.
\hfill $\Box$

Theorem~\ref{theorem_opt_p1p2_new} shows that problems (\ref{sub_problem1}) and (\ref{sub_problem2}) can be optimally solved by respectively checking the channel condition and assessing the values of ${\tilde R}$ under four feasible solutions.
Note that to make a distinction, we use $({\hat q}_1^*, {\hat q}_2^*)$ and $({\tilde q}_1^*, {\tilde q}_2^*)$ to respectively denote the optimal solutions of (\ref{sub_problem1}) and (\ref{sub_problem2}).
In the following theorem, we show that the optimal solution of (\ref{max_SR_new_jam2}), which is also the optimal solution of (\ref{max_SR_new_jam}), can be obtained based on Theorem~\ref{theorem_opt_p1p2_new}.
\begin{theorem}\label{theorem_opt_solu_new}
	The optimal solution of problem (\ref{max_SR_new_jam2}) can be obtained by considering 
	the following three different cases.
	\begin{itemize}
		\item If $\bm g_2^H \bm g_2 \leq \bm h_2^H \left( P_1 \bm h_1 \bm h_1^H + \bm I_{N_{\text b}} \right)^{-1} \bm h_2$, the optimal objective function value of (\ref{max_SR_new_jam2}) is ${\hat R} ({\hat q}_1^*, {\hat q}_2^*)$, where $({\hat q}_1^*, {\hat q}_2^*)$ is obtained by letting $({\hat P}_{1, {\text {lb}}}, {\hat P}_{1, {\text {ub}}}) = (0, P_1)$ and solving (\ref{sub_problem1}).
		\item If $\bm g_2^H \bm g_2 \geq \bm h_2^H \bm h_2$, the optimal objective function value of (\ref{max_SR_new_jam2}) is ${\tilde R} ({\tilde q}_1^*, {\tilde q}_2^*)$, where $({\tilde q}_1^*, {\tilde q}_2^*)$ is obtained by letting $({\tilde P}_{1, {\text {lb}}}, {\tilde P}_{1, {\text {ub}}}) = (0, P_1)$ and solving (\ref{sub_problem2}).
		\item If $\bm h_2^H \left( P_1 \bm h_1 \bm h_1^H + \bm I_{N_{\text b}} \right)^{-1} \bm h_2 < \bm g_2^H \bm g_2 < \bm h_2^H \bm h_2$, the optimal objective function value of (\ref{max_SR_new_jam2}) is 
		\begin{equation}\label{solu_case3}
		\max \{ {\hat R} ({\hat q}_1^*, {\hat q}_2^*), {\tilde R} ({\tilde q}_1^*, {\tilde q}_2^*) \},
		\end{equation}
		where $({\hat q}_1^*, {\hat q}_2^*)$ is obtained by letting $({\hat P}_{1, {\text {lb}}}, {\hat P}_{1, {\text {ub}}}) = (0, P_0')$ and solving (\ref{sub_problem1}), $({\tilde q}_1^*, {\tilde q}_2^*)$ is obtained by letting $({\tilde P}_{1, {\text {lb}}}, {\tilde P}_{1, {\text {ub}}}) = (P_0', P_1)$ and solving (\ref{sub_problem2}), and $P_0'$ is given in (\ref{P_0}).
	\end{itemize}
\end{theorem}
\itshape \textbf{Proof:} \upshape
See Appendix \ref{prove_theorem_opt_solu_new}.
\hfill $\Box$

\section{SD-based Low-complexity Precoder Design for the MIMO Case}
\label{SDLC_MIMO}

In this section, we consider the general MIMO system and propose specialized SD-based low-complexity (SDLC) schemes to address the precoder design problems for different jamming schemes.
Before introducing the schemes, we first present some preliminary results on the SD of matrices to set the stage for the subsequent discussions.

\subsection{Preliminaries on SD of Matrices}
\label{pre_SD}

\begin{lemma}\label{lemma_SD}
	\cite[Theorem~$2$]{au1971note} If two matrices $\bm F_1,~ \bm F_2 \in {\mathbb C}^{N \times N}$ are both positive semi-definite, they can be simultaneously diagonalized, i.e., there exists a matrix $\bm U$ such that $\bm U^H \bm F_1 \bm U$ and $\bm U^H \bm F_2 \bm U$ are diagonal matrices.
\end{lemma}
While \cite{au1971note} does not give an explicit way to compute the diagonalizing matrix $\bm U$, we provide a method in the following.
Denote the eigen-decomposition of $\bm F_1 + \bm F_2$ by 
\begin{equation}\label{eigende_HHGG}
\bm F_1 + \bm F_2 = \bm \varPsi_1 \begin{bmatrix}
\bm \varUpsilon & \bm 0 \\
\bm 0 & \bm 0 
\end{bmatrix} \bm \varPsi_1^H,
\end{equation}
where $\bm \varPsi_1 \in {\mathbb C}^{N \times N}$ is a unitary matrix, $\bm \varUpsilon \in {\mathbb R}^{{\hat N} \times {\hat N}}$ is a diagonal matrix with positive diagonal entries, and ${\hat N} = {\text {rank}} (\bm F_1 + \bm F_2)$.
Based on $\bm \varPsi_1$ and $\bm \varUpsilon$, we construct a matrix $\bm U_1$ as follows
\begin{equation}\label{U1}
\bm U_1 = \bm \varPsi_1 
\begin{bmatrix} 
\bm \varUpsilon^{- \frac{1}{2}} & \bm 0 \\ 
\bm 0 & \bm \varPi_1 
\end{bmatrix},
\end{equation}
where $\bm \varPi_1$ can be any square matrix of dimension $N - {\hat N}$.
Applying $\bm U_1$ to $\bm F_1 + \bm F_2$, it is obvious that
\begin{align}\label{UF1F2U}
\bm U_1^H (\bm F_1 + \bm F_2) \bm U_1 & = \begin{bmatrix}
\bm I_{{\hat N}} & \bm 0 \\
\bm 0 & \bm 0 
\end{bmatrix},
\end{align}
where the $({\hat N}+1)$-th to $N$-th diagonal entries are all zero.
Let $\bm u_{1,n}$ denote the $n$-th column of $\bm U_1$.
Then, we know from (\ref{UF1F2U}) that
\begin{equation}\label{all_zero1}
	\bm u_{1,n}^H (\bm F_1 + \bm F_2) \bm u_{1,n} = 0, ~\forall~ n \in \{{\hat N}+1, \cdots, N\},
\end{equation}
based on which we have 
\begin{equation}\label{all_zero2}
	\bm u_{1,n}^H \bm F_1 \bm u_{1,n} = 0, ~\forall~ n \in \{{\hat N}+1, \cdots, N\},
\end{equation}
since $\bm F_1$ and $\bm F_2$ are both positive semi-definite.
(\ref{all_zero2}) indicates that if we apply $\bm U_1$ to $\bm F_1$, the $({\hat N}+1)$-th to $N$-th diagonal entries of $\bm U_1^H \bm F_1 \bm U_1$ are all zero.
Then, it is known from the proof of \cite[Lemma~2]{au1971note} that $\bm U_1^H \bm F_1 \bm U_1$ takes on the following block matrix form
\begin{align}\label{UF1U}
	\bm U_1^H \bm F_1 \bm U_1 & = 
	\begin{bmatrix} 
		\bm J & \bm 0 \\
		\bm 0 & \bm 0 
	\end{bmatrix},
\end{align}
where $\bm J \in {\mathbb C}^{{\hat N} \times {\hat N}}$.
Denote the eigen-decomposition of $\bm J$ by
\begin{equation}\label{decomp_G0}
\bm J = \bm \varPsi_2 ~{\text {diag}} \{ \eta_1, \cdots, \eta_{{\hat N}} \}~ \bm \varPsi_2^H,
\end{equation}
where $\bm \varPsi_2 \in {\mathbb C}^{{\hat N} \times {\hat N}}$.
Since 
\begin{align}\label{UGGU}
	\bm U_1^H \bm F_1 \bm U_1 & = 
	\begin{bmatrix} 
		\bm J & \bm 0 \\
		\bm 0 & \bm 0 
	\end{bmatrix} \succeq \bm 0, \nonumber\\
	\bm U_1^H \bm F_2 \bm U_1 & = 
	\begin{bmatrix} 
		\bm I_{{\hat N}} - \bm J & \bm 0 \\
		\bm 0 & \bm 0 
	\end{bmatrix} \succeq \bm 0,
\end{align}
we have $ \bm 0 \preceq \bm J \preceq \bm I_{{\hat N}}$, which indicates that the eigenvalues in (\ref{decomp_G0}) satisfy $0 \leq \eta_t \leq 1, ~\forall~ n \in \{1, \cdots {\hat N}\} $.
We construct another matrix $\bm U_2$ and let $\bm U$ be the product of $\bm U_1$ and $\bm U_2$, 
\begin{align}\label{U2}
\bm U_2 & = 
\begin{bmatrix} 
\bm \varPsi_2 & \bm 0 \\
\bm 0 & \bm \varPi_2 
\end{bmatrix},\nonumber\\
\bm U & = \bm U_1 \bm U_2,
\end{align}
where $\bm \varPi_2$ can be any square matrix of dimension $N - {\hat N}$.
Then, it is known from (\ref{decomp_G0}), (\ref{UGGU}), and (\ref{U2}) that $\bm F_1$ and $\bm F_2$ can be simultaneously diagonalized by $\bm U$ as follows
\begin{align}\label{SDQ1Q2}
\bm U^H \bm F_1 \bm U & = {\text {diag}} \{ \eta_1, \cdots, \eta_{{\hat N}}, 0, \cdots, 0 \}, \nonumber\\
\bm U^H \bm F_2 \bm U & = {\text {diag}} \{ 1 - \eta_1, \cdots, 1 - \eta_{{\hat N}}, 0, \cdots, 0 \}.
\end{align}

Note that as shown above, if ${\hat N} < N$, $\bm \varPi_1$ and $\bm \varPi_2$ can be any square matrices of dimension $N - {\hat N}$.
Moreover, since the eigen-decomposition of a matrix is unique if and only if all its eigenvalues are distinct, $\bm \varPsi_1$ and $\bm \varPsi_2$ in (\ref{eigende_HHGG}) and (\ref{decomp_G0}) may not be unique.
Therefore, there may be different ways to simultaneously diagonalize $\bm F_1$ and $\bm F_2$.

\subsection{GN Scheme}
\label{GN_jam_MIMO}

If Tx2 adopts the GN scheme for cooperative jamming, we consider problem (\ref{max_SR_GN}). 
A similar problem has been studied in \cite{fakoorian2011solutions} under the 
constraint $\bm Q_1 \preceq \bm S$ (see \cite[(4)]{fakoorian2011solutions}) for Tx1, 
where $\bm S$ is some positive semi-definite matrix.
Obviously, the constraint ${\text {tr}}(\bm Q_1) \leq P_1$ considered in this paper is more general.
Due to the more restrictive constraint, the system's performance may be limited and the scheme provided in \cite{fakoorian2011solutions} does not apply here.

Since (\ref{max_SR_GN}) is non-convex and analytically intractable, we provide a heuristic solution by iteratively optimizing $\bm Q_1$ and $\bm Q_2$, i.e., dealing with the following problems in an alternative manner
\begin{align}
	\mathop {\max }\limits_{ \bm Q_1 } \quad & \ln | \bm H_1 \bm Q_1 \bm H_1^H \left( \bm C_2 + \bm I_{N_{\text b}} \right)^{-1} + \bm I_{N_{\text b}} | \nonumber\\
	- & \ln | \bm G_1 \bm Q_1 \bm G_1^H \left( \bm D_2 + \bm I_{N_{\text e}} \right)^{-1} + \bm I_{N_{\text e}} | \nonumber\\
	\text{s.t.} \quad\; & \bm Q_1 \succeq \bm 0,~ {\text {tr}}(\bm Q_1) \leq P_1, \label{max_SR_GN_jam_F1}
	\\
	\mathop {\max }\limits_{ \bm Q_2 } \quad & \ln | \bm C_1 \left( \bm H_2 \bm Q_2 \bm H_2^H + \bm I_{N_{\text b}} \right)^{-1} + \bm I_{N_{\text b}} | \nonumber\\ 
	- & \ln | \bm D_1 \left( \bm G_2 \bm Q_2 \bm G_2^H + \bm I_{N_{\text e}} \right)^{-1} + \bm I_{N_{\text e}} | \nonumber\\
	\text{s.t.} \quad\; & \bm Q_2 \succeq \bm 0,~ {\text {tr}}(\bm Q_2) \leq P_2, \label{max_SR_GN_jam_F2}
\end{align}	
where $\bm C_k = \bm H_k \bm Q_k \bm H_k^H$, $\bm D_k = \bm G_k \bm Q_k \bm G_k^H, \forall k \in \{ 1, 2 \}$, and $[\cdot]^+$ is omitted for convenience.
In the following we deal with (\ref{max_SR_GN_jam_F1}) and (\ref{max_SR_GN_jam_F2})  using the SD technique, and show that a good feasible point in closed form can be obtained for each of them.

\subsubsection{SD-based scheme for solving (\ref{max_SR_GN_jam_F1})}

Denote ${\hat{ \bm H}}_1 = \left( \bm C_2 + \bm I_{N_{\text b}} \right)^{-\frac{1}{2}} \bm H_1$ and ${\hat{ \bm G}}_1 = \left( \bm D_2 + \bm I_{N_{\text e}} \right)^{-\frac{1}{2}} \bm G_1$.
Then, using (\ref{O1O2_1}), problem (\ref{max_SR_GN_jam_F1}) can be equivalently transformed to
\begin{align}\label{max_SR_GN_jam_F1_v2}
	\mathop {\max }\limits_{ \bm Q_1 } \quad & \ln | {\hat{ \bm H}}_1 \bm Q_1 {\hat{ \bm H}}_1^H + \bm I_{N_{\text b}} | - \ln | {\hat{ \bm G}}_1 \bm Q_1 {\hat{ \bm G}}_1^H + \bm I_{N_{\text e}} | \nonumber\\
	\text{s.t.} \quad\; & \bm Q_1 \succeq \bm 0,~ {\text {tr}}(\bm Q_1) \leq P_1.
\end{align}
Obviously, (\ref{max_SR_GN_jam_F1_v2}) can be seen as the secrecy rate maximization problem of the classical MIMOME channel \cite{khisti2010secure, zhang2020rotation}, where ${\hat{ \bm H}}_1$ and ${\hat{ \bm G}}_1$ are respectively the channel matrices from the transmitter to Bob and Eve.
Such a problem has been widely studied and the analytical capacity-achieving solution exists for some special cases \cite{vaezi2017optimal, vaezi2017mimo, zhang2020rotation, xu2022sd}.
However, the analytical solution for the general MIMO channel is still an open problem.
In \cite{xu2022sd}, the SD technique was proposed to solve the problem, and its superiority in terms of both secrecy rate and computational complexity, over the iterative MM-based algorithm \cite{xu2022achievable} as well as the generalized singular value decomposition (GSVD) approach \cite{fakoorian2012optimal}, was verified.
For completeness, we provide the main steps of this scheme for solving (\ref{max_SR_GN_jam_F1_v2}) below.

Since ${\hat {\bm H}}_1^H {\hat {\bm H}}_1$ and ${\hat{ \bm G}}_1^H {\hat{ \bm G}}_1$ are both positive semi-definite, we known from Lemma~\ref{lemma_SD} that they can be simultaneously diagonalized.
A matrix $\bm V$ can be constructed such that
\begin{align}\label{SDHHGG1}
	\bm V^H {\hat {\bm H}}_1^H {\hat {\bm H}}_1 \bm V & = {\text {diag}} \{ \phi_1, \cdots, \phi_{{\hat N}_1}, 0, \cdots, 0 \}, \nonumber\\
	\bm V^H {\hat{ \bm G}}_1^H {\hat{ \bm G}}_1 \bm V & = {\text {diag}} \{ 1 \!-\! \phi_1, \cdots, 1 \!-\! \phi_{{\hat N}_1}, 0, \cdots, 0 \},
\end{align}
where ${\hat N}_1 = {\text {rank}} ({\hat {\bm H}}_1^H {\hat {\bm H}}_1 + {\hat{ \bm G}}_1^H {\hat{ \bm G}}_1)$ and $0 \leq \phi_n \leq 1, ~\forall~ n \in \{1, \cdots, {\hat N}_1\} $.
Let 
\begin{equation}\label{F1_decomp}
	\bm Q_1 = \bm V \bm \varTheta \bm V^H,
\end{equation}
where $\bm \varTheta \triangleq {\text {diag}} \{ \theta_1, \cdots, \theta_{N_1} \}$ has non-negative real diagonal entries.
Based on (\ref{O1O2_1}), (\ref{O1O2_3}), (\ref{SDHHGG1}) and (\ref{F1_decomp}), the objective function of (\ref{max_SR_GN_jam_F1_v2}) and ${\text {tr}}(\bm Q_1)$ in the constraint can be rewritten as
\begin{align}
	& \ln | {\hat{ \bm H}}_1 \bm Q_1 {\hat{ \bm H}}_1^H + \bm I_{N_{\text b}} | - \ln | {\hat{ \bm G}}_1 \bm Q_1 {\hat{ \bm G}}_1^H + \bm I_{N_{\text e}} | \nonumber\\
	= & \ln | \bm V^H {\hat{ \bm H}}_1^H {\hat{ \bm H}}_1 \bm V \bm \varTheta + \bm I_{N_{\text b}} | - \ln | \bm V^H {\hat{ \bm G}}_1^H {\hat{ \bm G}}_1 \bm V \bm \varTheta + \bm I_{N_{\text e}} | \nonumber\\
	= & \sum_{n=1}^{{\hat N}_1} \left[ \ln (\phi_n \theta_n + 1) - \ln ((1 - \phi_n) \theta_n + 1) \right], \label{SD_ob1}
\end{align}
and
\begin{align}
	{\text {tr}}(\bm Q_1) & = {\text {tr}}( \bm V^H \bm V \bm \varTheta) \nonumber\\
	& = \sum_{n=1}^{N_1} \left\| \bm v_n \right\|^2 \theta_n, \label{diag_trace_F1}
\end{align}
where $\bm v_n$ is the $n$-th column of $\bm V$.
Then, instead of directly solving (\ref{max_SR_GN_jam_F1_v2}), we consider the following problem
\begin{subequations}\label{optimize_theta}
	\begin{align}
		\mathop {\min }\limits_{\bm \varTheta} \quad & \sum_{n=1}^{{\hat N}_1} \left[ - \ln (\phi_n \theta_n + 1) + \ln ((1 - \phi_n) \theta_n + 1) \right] \label{optimize_theta_a}\\
		\text{s.t.} \quad\; &  \theta_n \geq 0, ~\forall~ n \in \{ 1, \cdots, N_1 \}, \label{optimize_theta_b}\\
		& \sum_{n=1}^{N_1} \left\| \bm v_n \right\|^2 \theta_n \leq P_2. \label{optimize_theta_c}
	\end{align}
\end{subequations}
Although (\ref{optimize_theta}) is non-convex, its optimal solution can be obtained in closed form  \cite[Theorem~$1$]{xu2022sd}.
The only additional step required is a one-dimensional line search with respect to a Lagrange multiplier variable.
\begin{lemma}\label{theorem_optimal_a}
	The optimal solution of problem (\ref{optimize_theta}) is
	\begin{equation}\label{optimal_A}
		\theta_n^* (\!\alpha^*\!) \!=\! \left\{\!\!\!\!
		\begin{array}{ll}
			0, ~{\text {if}}~ {\hat N}_1 \!\leq\! n \!\leq\! N_1 ~{\text {or}}~ \!\left\{\! 1 \!\leq\! n \!\leq\! {\hat N}_1 ~{\text {and}}~ 0 \!\leq\! \phi_n \!\leq\! \frac{1}{2} \!\right\}\!, \vspace{0.3em} \\
			\left[ \frac{1}{\alpha^* \left\| \bm v_n \right\|^2} -1 \right]^+, ~{\text {if}}~ 1 \leq n \leq {\hat N}_1 ~{\text {and}}~ \phi_n = 1, \vspace{0.3em} \\
			\frac{\left[ - 1 + \sqrt{1 - 4 \phi_n (1 - \phi_n) \left( 1 + \frac{1 - 2 \phi_n}{\alpha^* \left\| \bm v_n \right\|^2} \right)} \right]^{+}}{2 \phi_n (1 - \phi_n)}, {\text {if}}~ 1 \!\leq\! n \!\leq\! {\hat N}_1\\
			\quad\quad\quad\quad\quad\quad\quad\quad\quad\quad\quad\quad\quad {\text {and}}~ \frac{1}{2} < \phi_n < 1,
		\end{array} \right.
	\end{equation}
	where $\alpha^*$ is the Lagrange multiplier attached to the constraint (\ref{optimize_theta_c}), and it can be found by using the bisection searching method such that (\ref{optimize_theta_c}) holds with equality.
	\hfill $\Box$
\end{lemma}

Once problem (\ref{optimize_theta}) is solved, a solution to (\ref{max_SR_GN_jam_F1}) and (\ref{max_SR_GN_jam_F1_v2}) can be obtained based on (\ref{F1_decomp}).
Denote the SD-based low-complexity approach proposed for solving (\ref{max_SR_GN_jam_F1}) by {\textbf {SDLC1}}.
Note that due to the limitations imposed by (\ref{F1_decomp}) on the formation of $\bm Q_1$, while (\ref{optimize_theta}) can be solved optimally, the solution to (\ref{max_SR_GN_jam_F1}) obtained from Theorem~\ref{theorem_optimal_a} and (\ref{F1_decomp}) may not necessarily be optimal.

\subsubsection{SD-based scheme for solving (\ref{max_SR_GN_jam_F2})}
Since $\bm Q_2$ appears in the inverse term of the objective function, the SD technique cannot be directly applied to deal with (\ref{max_SR_GN_jam_F2}).
In the following we first make a transformation to its objective function, and then show that the SD technique can still be applied and a good feasible point in closed form can be obtained.
Denote ${\hat{ \bm H}}_2 = \left( \bm C_1 + \bm I_{N_{\text b}} \right)^{-\frac{1}{2}} \bm H_2$ and ${\hat{ \bm G}}_2 = \left( \bm D_1 + \bm I_{N_{\text e}} \right)^{-\frac{1}{2}} \bm G_2$.
Then, based on (\ref{O1O2_1}) and (\ref{O1O2_2}), the objective function of (\ref{max_SR_GN_jam_F2}) can be rewritten as
\begin{align}\label{first_logdet}
	& \ln | \bm H_2 \bm Q_2 \bm H_2^H + \bm C_1 + \bm I_{N_{\text b}} | - \ln | \bm H_2 \bm Q_2 \bm H_2^H + \bm I_{N_{\text b}} | \nonumber\\
	- & \ln | \bm G_2 \bm Q_2 \bm G_2^H + \bm D_1 + \bm I_{N_{\text e}} | + \ln | \bm G_2 \bm Q_2 \bm G_2^H + \bm I_{N_{\text e}} | \nonumber\\
	= & \ln | {\hat{ \bm H}}_2 \bm Q_2 {\hat{ \bm H}}_2^H + \bm I_{N_{\text b}} | - \ln | \bm H_2 \bm Q_2 \bm H_2^H + \bm I_{N_{\text b}} | \nonumber\\
	- & \ln | {\hat{ \bm G}}_2 \bm Q_2 {\hat{ \bm G}}_2^H + \bm I_{N_{\text e}} | + \ln | \bm G_2 \bm Q_2 \bm G_2^H + \bm I_{N_{\text e}} | \nonumber\\
	+ & \ln | \bm C_1 + \bm I_{N_{\text b}} | - \ln | \bm D_1 + \bm I_{N_{\text e}} |.
\end{align}
Neglecting the constant terms in (\ref{first_logdet}), problem (\ref{max_SR_GN_jam_F2}) can be equivalently transformed to 
\begin{subequations}\label{max_SR_GN_jam_F2_2}
\begin{align}
	\mathop {\min }\limits_{ \bm Q_2 }  \; & - \ln | {\hat{ \bm H}}_2 \bm Q_2 {\hat{ \bm H}}_2^H + \bm I_{N_{\text b}} | - \ln | \bm G_2 \bm Q_2 \bm G_2^H + \bm I_{N_{\text e}} | \nonumber\\
	\; & + \ln | \bm H_2 \bm Q_2 \bm H_2^H + \bm I_{N_{\text b}} | + \ln | {\hat{ \bm G}}_2 \bm Q_2 {\hat{ \bm G}}_2^H + \bm I_{N_{\text e}} | \label{max_SR_GN_jam_F2_2a}\\
	\text{s.t.} \;\; & \bm Q_2 \succeq \bm 0,~ {\text {tr}}(\bm Q_2) \leq P_2. \label{max_SR_GN_jam_F2_2b}
\end{align}
\end{subequations}

Due to the log-concavity of determinant, (\ref{max_SR_GN_jam_F2_2}) is a difference of convex (DC) programming.
The iterative MM-based method can thus be applied to get a sequence of convex subproblems.
Specifically, let $\bm Q_{2, 0}$ denote the solution obtained in the previous iteration.
Using the first-order Taylor series approximation to linearize the third and fourth log-determinant terms in (\ref{max_SR_GN_jam_F2_2a}) \cite{deadman2016taylor}, we get the following upper bound
\begin{align}\label{Taylor_appro}
	& \ln | \bm H_2 \bm Q_2 \bm H_2^H + \bm I_{N_{\text b}} | + \ln | {\hat{ \bm G}}_2 \bm Q_2 {\hat{ \bm G}}_2^H + \bm I_{N_{\text e}} | \nonumber\\
	\leq & \ln | \bm H_2 \bm Q_{2, 0} \bm H_2^H + \bm I_{N_{\text b}} | + \ln | {\hat{ \bm G}}_2 \bm Q_{2, 0} {\hat{ \bm G}}_2^H + \bm I_{N_{\text e}} | \nonumber\\
	+ & {\text {tr}} \left( \bm A \bm Q_2 \right) - {\text {tr}} \left( \bm A \bm Q_{2, 0} \right),
\end{align}
where $\bm A \!=\! \bm H_2^H \!\left( \bm H_2 \bm Q_{2, 0} \bm H_2^H \!+\! \bm I_{N_{\text b}} \right)^{-1}\!\! \bm H_2 + {\hat{ \bm G}}_2^H ( {\hat{ \bm G}}_2 \bm Q_{2, 0} {\hat{ \bm G}}_2^H $ $ + \bm I_{N_{\text e}} )^{-1} {\hat{ \bm G}}_2$.
Then, based on (\ref{Taylor_appro}), we solve (\ref{max_SR_GN_jam_F2_2}) by iteratively dealing with the following problem
\begin{subequations}\label{R1_s_k2_max2}
	\begin{align}
	\mathop {\min }\limits_{\bm Q_2} \quad & - \ln | {\hat{ \bm H}}_2 \bm Q_2 {\hat{ \bm H}}_2^H + \bm I_{N_{\text b}} | - \ln | \bm G_2 \bm Q_2 \bm G_2^H + \bm I_{N_{\text e}} | \nonumber\\
	& + {\text {tr}} \left( \bm A \bm Q_2 \right) \label{R1_s_k2_max2_a}\\
	\text{s.t.} \quad\, &  \bm Q_2 \succeq \bm 0,~ {\text {tr}}(\bm Q_2) \leq P_2, \label{R1_s_k2_max2_b}
	\end{align}
\end{subequations}
which is convex and can thus be optimally solved using some standard tools like the interior-point method.
However, the presence of log-determinant terms in the objective function makes it extremely time-consuming \cite{xu2022sd}.
Hence, we propose to solve it using again the SD technique.

Since ${\hat {\bm H}}_2^H {\hat {\bm H}}_2$ and $\bm G_2^H \bm G_2$ are both positive semi-definite, according to Lemma~\ref{lemma_SD}, a diagonalizing matrix $\bm W$ can be constructed such that 
\begin{align}\label{SDHHGG2}
\bm W^H {\hat {\bm H}}_2^H {\hat {\bm H}}_2 \bm W & = {\text {diag}} \{ \rho_1, \cdots, \rho_{{\hat N}_2}, 0, \cdots, 0 \}, \nonumber\\
\bm W^H \bm G_2^H \bm G_2 \bm W & = {\text {diag}} \{ 1 \!-\! \rho_1, \cdots, 1 \!-\! \rho_{{\hat N}_2}, 0, \cdots, 0 \},
\end{align}
where ${\hat N}_2 = {\text {rank}} ({\hat {\bm H}}_2^H {\hat {\bm H}}_2 + \bm G_2^H \bm G_2)$ and $0 \leq \rho_n \leq 1, ~\forall~ n \in \{1, \cdots, {\hat N}_2\} $.
Let 
\begin{equation}\label{F2_decomp}
\bm Q_2 = \bm W \bm \varLambda \bm W^H,
\end{equation}
where $\bm \varLambda \triangleq {\text {diag}} \{ \lambda_1, \cdots, \lambda_{N_2} \}$ has non-negative real diagonal entries.
Based on (\ref{O1O2_1}), (\ref{O1O2_3}), (\ref{SDHHGG2}), and (\ref{F2_decomp}), the objective function (\ref{R1_s_k2_max2_a}) and ${\text {tr}}(\bm Q_2)$ in constraint (\ref{R1_s_k2_max2_b}) can be rewritten as
\begin{align}
& - \ln | {\hat{ \bm H}}_2 \bm Q_2 {\hat{ \bm H}}_2^H + \bm I_{N_{\text b}} | - \ln | \bm G_2 \bm Q_2 \bm G_2^H + \bm I_{N_{\text e}} | + {\text {tr}} \left( \bm A \bm Q_2 \right) \nonumber\\
& = -\! \ln\! | \bm W^H {\hat{ \bm H}}_2^H {\hat{ \bm H}}_2 \bm W \bm \varLambda \!+\! \bm I_{N_{\text b}} | \!-\! \ln\! | \bm W^H \bm G_2^H \bm G_2 \bm W \bm \varLambda \!+\! \bm I_{N_{\text e}} | \nonumber\\
& + {\text {tr}} \left( \bm W^H \bm A \bm W \bm \varLambda \right) \nonumber\\
& = \sum_{n=1}^{{\hat N}_2} \left[ - \ln (\rho_n \lambda_n \!+\! 1) \!-\! \ln ((1 \!-\! \rho_n) \lambda_n \!+\! 1) \right] \!+\! \sum_{n=1}^{N_2} a_n \lambda_n,\!\!\! \label{SD_ob2}
\end{align}
and
\begin{align}
{\text {tr}}(\bm Q_2) & = {\text {tr}}( \bm W^H \bm W \bm \varLambda) \nonumber\\
& = \sum_{n=1}^{N_2} \left\| \bm w_n \right\|^2 \lambda_n, \label{diag_trace_F2}
\end{align}
where $a_n$ is the $n$-th diagonal entry of $\bm W^H \bm A \bm W$ and $\bm w_n$ is the $n$-th column of $\bm W$.
Then, instead of solving (\ref{R1_s_k2_max2}) using the general tools, we consider the following problem
\begin{subequations}\label{SD_problem3}
	\begin{align}
	\!\!\!\!\mathop {\min }\limits_{\bm \varLambda} \; & \sum_{n=1}^{{\hat N}_2} \!\left[ -\! \ln (\rho_n \lambda_n \!\!+\!\! 1) \!-\! \ln ((1 \!-\! \rho_n) \lambda_n \!+\! 1) \right] \!+\! \sum_{n=1}^{N_2} \!a_n \lambda_n\!\!\!  \label{SD_problem3_a}\\
	\!\!\!\!\text{s.t.} \;\; &  \lambda_n \geq 0, ~\forall~ n \in \{ 1, \cdots, N_2 \}, \label{SD_problem3_b}\\
	& \sum_{n=1}^{N_2} \left\| \bm w_n \right\|^2 \lambda_n \leq P_2, \label{SD_problem3_c}
	\end{align}
\end{subequations}
which is convex.
In the following theorem we provide its optimal solution in closed form (the only additional step required is a one-dimensional line search).
\begin{theorem}\label{theo_opt_lambda}
	The optimal solution of problem (\ref{SD_problem3}) is
	\begin{equation}\label{optimal_lambda}
	\lambda_n^* (\beta^*) \!=\! \left\{\!\!\!
	\begin{array}{ll}
	0, ~{\text {if}}~ {\hat N}_2 + 1 \leq n \leq N_2, \vspace{0.3em} \\
	\left[ \frac{1}{a_n \!+\! \beta^* \left\| \bm w_n \right\|^2} \!-\! 1 \right]^{\!+}\!\!, ~{\text {if}}~ 1 \!\leq\! n \!\leq\! {\hat N}_2 ~{\text {and}}~ \rho_n \!=\! 0 \!~{\text {or}}~\! 1, \vspace{0.3em} \\
	\frac{\left[ \!-\! 1 \!+\! \frac{2 \rho_n (1 - \rho_n)}{a_n \!+\! \beta^* \left\| \bm w_n \right\|^2} \!+\! \sqrt{ (2 \rho_n \!-\! 1)^2 \!+\! \frac{4 \rho_n^2 (1 - \rho_n)^2}{\left( a_n \!+\! \beta^* \left\| \bm w_n \right\|^2 \right)^2}} \right]^+}{2 \rho_n (1 - \rho_n)},\\
	\quad\quad\quad\quad\quad\quad\quad {\text {if}}~ 1 \leq n \leq {\hat N}_2 ~{\text {and}}~ 0 < \rho_n < 1.
	\end{array} \right.
	\end{equation}
	If $\lambda_n^* (0)$ makes the constraint (\ref{SD_problem3_c}) hold, $\beta^* = 0$.
	Otherwise, $\beta^*$ can be found using the bisection searching method such that (\ref{SD_problem3_c}) holds with equality.
\end{theorem}
\itshape \textbf{Proof:}  \upshape
See Appendix~\ref{prove_theo_opt_lambda}.
\hfill $\Box$

To distinguish from SDLC1, we denote the SD-based low-complexity scheme for solving (\ref{R1_s_k2_max2}) by {\textbf {SDLC2}}.
Then, the proposed approach to solve (\ref{max_SR_GN})
consists of alternatively applying SDLC1 and SDLC2 for a specific fine number of iterations, 
as summarized in Algorithm~\ref{algorithm1}.

\begin{algorithm}[h]
	\begin{algorithmic}[1]
		\caption{SD-based low-complexity scheme for solving (\ref{max_SR_GN})}
		\State Initialize ${\bm Q}_1$ and $\bm Q_2$.
		\For{$l_1 = 1:L_1$}
		\State Solve (\ref{max_SR_GN_jam_F1}) and update $\bm Q_1$ using SDLC1.
		\For{$l_2 = 1:L_2$}
		\State Let ${\bm Q}_{2, 0} = \bm Q_2$ and calculate $\bm A$. Solve (\ref{R1_s_k2_max2}) and update $\bm Q_2$ using SDLC2.
		\EndFor
		\EndFor
		\label{algorithm1}
	\end{algorithmic}
\end{algorithm}

\subsection{EJ Scheme}
\label{IT_CJ_MIMO}

Now we consider the EJ scheme for Tx2 and problem (\ref{max_SR_EJ}).
Since $R_{\text {EJ}} = \max \{ \min \{ {\hat R}, {\tilde R} \}, {\bar R} \}$, (\ref{max_SR_EJ}) can be decomposed into two subproblems as follows
\begin{align}
	\mathop {\max }\limits_{\bm Q_1} \quad & {\bar R} \nonumber\\
	\text{s.t.} \quad\; & \bm Q_1 \succeq \bm 0,~ {\text {tr}}(\bm Q_1) \leq P_1,\label{max_SR_bar_MIMO}\\
	\mathop {\max }\limits_{ \bm Q_1, \bm Q_2 } \quad & \min \{ {\hat R}, {\tilde R} \} \nonumber\\
	\text{s.t.} \quad\; & \bm Q_k \succeq \bm 0,~ {\text {tr}}(\bm Q_k) \leq P_k, ~\forall~ k \in \{ 1, 2 \}. \label{max_SR_hat_tilde_MIMO}
\end{align}
Let $\bm Q_1^\star$ and $(\bm Q_1^*, \bm Q_2^*)$ respectively denote the optimal solutions of (\ref{max_SR_bar_MIMO}) and (\ref{max_SR_hat_tilde_MIMO}).
From the expressions of ${\bar R}$, ${\hat R}$, and ${\tilde R}$ in (\ref{SR_hat_tilde_bar_MIMO}) we know that 
\begin{align}\label{ineq1}
	\min \{ {\hat R} (\bm Q_1^*, \bm Q_2^*), {\tilde R} (\bm Q_1^*, \bm Q_2^*) \} & \geq \min \{ {\hat R} (\bm Q_1^\star, \bm 0), {\tilde R} (\bm Q_1^\star, \bm 0) \} \nonumber\\
	& = {\bar R} (\bm Q_1^\star),
\end{align}
implying that the optimal solution $(\bm Q_1^*, \bm Q_2^*)$ of (\ref{max_SR_hat_tilde_MIMO}) is the optimal solution of (\ref{max_SR_EJ}), and
\begin{equation}\label{ineq2}
	R_{\text {EJ}} (\bm Q_1^*, \bm Q_2^*) = \min \{ {\hat R} (\bm Q_1^*, \bm Q_2^*), {\tilde R} (\bm Q_1^*, \bm Q_2^*) \}.
\end{equation}
Therefore, if (\ref{max_SR_hat_tilde_MIMO}) can be optimally solved, there is actually no need to consider (\ref{max_SR_bar_MIMO}).
However, unlike the SIMO case, this is no longer possible.
Therefore, we solve (\ref{max_SR_bar_MIMO}) and (\ref{max_SR_hat_tilde_MIMO}) separately, and obtain an achievable lower bound for (\ref{max_SR_EJ}).

Since ${\bar R}$ takes on the same form as the objective function of (\ref{max_SR_GN_jam_F1_v2}), problem (\ref{max_SR_bar_MIMO}) can be efficiently solved by SDLC1.
Next, we solve (\ref{max_SR_hat_tilde_MIMO}) by separately maximizing ${\hat R}$ and ${\tilde R}$
\begin{align}
	\mathop {\max }\limits_{\bm Q_1, \bm Q_2} \quad & {\hat R} \nonumber\\
	\text{s.t.} \quad\; & \bm Q_k \succeq \bm 0,~ {\text {tr}}(\bm Q_k) \leq P_k, ~\forall~ k \in \{ 1, 2 \}, \label{max_SR_hat_MIMO}\\
	\mathop {\max }\limits_{\bm Q_1, \bm Q_2} \quad & {\tilde R} \nonumber\\
	\text{s.t.} \quad\; & \bm Q_k \succeq \bm 0,~ {\text {tr}}(\bm Q_k) \leq P_k, ~\forall~ k \in \{ 1, 2 \}. \label{max_SR_tilde_MIMO}
\end{align}
Let ${\bar {\bm Q}}_1$, $({\hat {\bm Q}}_1, {\hat {\bm Q}}_2)$, and $({\tilde {\bm Q}}_1, {\tilde {\bm Q}}_2)$ respectively denote the solutions (not necessarily optimal) of (\ref{max_SR_bar_MIMO}), (\ref{max_SR_hat_MIMO}), and (\ref{max_SR_tilde_MIMO}).
Since $(\bm Q_1^*, \bm Q_2^*)$ is the optimal solution of (\ref{max_SR_EJ}), we must have
\begin{equation}\label{rate_lb}
	R_{\text {EJ}} (\bm Q_1^*, \bm Q_2^*) \!\geq\! \max \{\! R_{\text {EJ}} ({\bar {\bm Q}}_1, \bm 0), R_{\text {EJ}} ({\hat {\bm Q}}_1, {\hat {\bm Q}}_2), R_{\text {EJ}} ({\tilde {\bm Q}}_1, {\tilde {\bm Q}}_2) \!\}.
\end{equation}
According to (\ref{rate_lb}), an achievable lower bound to $R_{\text {EJ}} (\bm Q_1^*, \bm Q_2^*)$ can be obtained by separately solving (\ref{max_SR_bar_MIMO}), (\ref{max_SR_hat_MIMO}), and (\ref{max_SR_tilde_MIMO}).
Once they are solved, we choose the point among $({\bar {\bm Q}}_1, \bm 0)$, $({\hat {\bm Q}}_1, {\hat {\bm Q}}_2)$, and $({\tilde {\bm Q}}_1, {\tilde {\bm Q}}_2)$ that produces the maximum $R_{\text {EJ}} (\bm Q_1, \bm Q_2)$, as the heuristic 
solution to (\ref{max_SR_EJ}).

As stated above, (\ref{max_SR_bar_MIMO}) can be solved by SDLC1.
Now we show that (\ref{max_SR_hat_MIMO}) and (\ref{max_SR_tilde_MIMO}) can also be solved by iteratively applying SDLC1.
Since they are both non-convex, we iteratively optimize $\bm Q_1$ and $\bm Q_2$.
For problem (\ref{max_SR_hat_MIMO}), by respectively fixing $\bm Q_2$ and $\bm Q_1$, we consider subproblems
\begin{align}
	\mathop {\max }\limits_{ \bm Q_1 } \; & \ln\! | \bm H_1 \bm Q_1 \bm H_1^H \!\!+\! \bm I_{N_{\text b}} | \!-\! \ln\! | \bm G_1 \bm Q_1 \bm G_1^H \!\left( \bm D_2 \!+\! \bm I_{N_{\text e}} \right)^{\!-1} \!\!+\! \bm I_{N_{\text e}} | \nonumber\\
	\text{s.t.} \;\; & \bm Q_1 \succeq \bm 0,~ {\text {tr}}(\bm Q_1) \leq P_1,\label{max_SR_hat_F1}\\
	\mathop {\max }\limits_{ \bm Q_2 } \; & \ln\! | \bm G_2 \bm Q_2 \bm G_2^H \!\!+\! \bm I_{N_{\text e}} | \!-\! \ln\! | \bm G_2 \bm Q_2 \bm G_2^H \!( \bm D_1 \!+\! \bm I_{N_{\text e}} )^{\!-1}\!\! +\! \bm I_{N_{\text e}} | \nonumber\\
	\text{s.t.} \;\; & \bm Q_2 \succeq \bm 0,~ {\text {tr}}(\bm Q_2) \leq P_2, \label{max_SR_hat_F2}
\end{align}
where $\bm D_1$ and $\bm D_2$ are defined in (\ref{max_SR_GN_jam_F2}).
Obviously, (\ref{max_SR_hat_F1}) and (\ref{max_SR_hat_F2}) have similar expressions as (\ref{max_SR_GN_jam_F1}).
Therefore, both of them can be efficiently solved by using SDLC1.
Then, we solve (\ref{max_SR_tilde_MIMO}) also in an alternative manner.
With fixed $\bm Q_2$ and $\bm Q_1$, we get two subproblems as follows
\begin{align}
	\mathop {\max }\limits_{ \bm Q_1 } \quad & \ln | \bm H_1 \bm Q_1 \bm H_1^H \left( \bm C_2 + \bm I_{N_{\text b}} \right)^{-1} + \bm I_{N_{\text b}} | \nonumber\\
	- & \ln | \bm G_1 \bm Q_1 \bm G_1^H \left( \bm D_2 + \bm I_{N_{\text e}} \right)^{-1} + \bm I_{N_{\text e}} | \nonumber\\
	\text{s.t.} \quad\; & \bm Q_1 \succeq \bm 0,~ {\text {tr}}(\bm Q_1) \leq P_1,\label{max_SR_tilde_F1}\\
	\mathop {\max }\limits_{ \bm Q_2 } \quad & \ln | \bm H_2 \bm Q_2 \bm H_2^H ( \bm C_1 + \bm I_{N_{\text b}} )^{-1} + \bm I_{N_{\text b}} | \nonumber\\
	- & \ln | \bm G_2 \bm Q_2 \bm G_2^H ( \bm D_1 + \bm I_{N_{\text e}} )^{-1} + \bm I_{N_{\text e}} | \nonumber\\
	\text{s.t.} \quad\; & \bm Q_2 \succeq \bm 0,~ {\text {tr}}(\bm Q_2) \leq P_2.\label{max_SR_tilde_F2}
\end{align}
where $\bm C_1$ and $\bm C_2$ are defined in (\ref{max_SR_GN_jam_F2}).
Obviously, (\ref{max_SR_tilde_F1}) and (\ref{max_SR_tilde_F2}) can also be solved by SDLC1.

The method for solving (\ref{max_SR_EJ}) is summarized in the following Algorithm~\ref{algorithm2}.

\begin{algorithm}[h]
	\begin{algorithmic}[1]
		\caption{SD-based low-complexity scheme for solving (\ref{max_SR_EJ})}
		\State Solve (\ref{max_SR_bar_MIMO}) using SDLC1 and obtain ${\bar {\bm Q}}_1$.
		\State Initialize ${\bm Q}_1$ and $\bm Q_2$.
		\For{$l_3 = 1:L_3$}
		\State Update $\bm Q_1$ and $\bm Q_2$ by respectively solving (\ref{max_SR_hat_F1}) and (\ref{max_SR_hat_F2}) using SDLC1.
		\EndFor
		\State Let $({\hat {\bm Q}}_1, {\hat {\bm Q}}_2) = (\bm Q_1, \bm Q_2)$ be the solution to (\ref{max_SR_hat_MIMO}).
		\State Re-initialize ${\bm Q}_1$ and $\bm Q_2$.
		\For{$l_4 = 1:L_4$}
		\State Update $\bm Q_1$ and $\bm Q_2$ by respectively solving (\ref{max_SR_tilde_F1}) and (\ref{max_SR_tilde_F2}) using SDLC1.
		\EndFor
		\State Let $({\tilde {\bm Q}}_1, {\tilde {\bm Q}}_2) = (\bm Q_1, \bm Q_2)$ be the solution to (\ref{max_SR_tilde_MIMO}).
		\State Choose the point among $({\bar {\bm Q}}_1, \bm 0)$, $({\hat {\bm Q}}_1, {\hat {\bm Q}}_2)$, and $({\tilde {\bm Q}}_1, {\tilde {\bm Q}}_2)$ that maximizes $R_{\text {EJ}} (\bm Q_1, \bm Q_2)$, as the solution to (\ref{max_SR_EJ}).
		\label{algorithm2}
	\end{algorithmic}
\end{algorithm}

\subsection{Convergence and Complexity Analysis}
\label{convergence_complexity}
\subsubsection{Convergence Analysis}
\label{convergence_analysis}
As shown in Algorithm~\ref{algorithm1} and Algorithm~\ref{algorithm2}, we execute the proposed SDLC1 or SDLC2 schemes for a fixed number of iterations in both algorithms.
\subsubsection{Complexity Analysis}
\label{complexity_analysis}
To evaluate the complexity of the proposed algorithms, we count the total number of floating-point operations (FLOPs), where one FLOP represents a complex multiplication or summation, express it as a polynomial function of the dimensions of the matrices involved, and simplify the expression by ignoring all terms except the leading (i.e., highest order or dominant) terms \cite{boyd2004convex, hunger2005floating}.
It is worth mentioning that the given analysis only shows how the bounds on computational complexity are related to different problem dimensions. 
The actual load may vary depending on the structure simplifications and used numerical solvers.

For convenience, we assume equal number of antennas for both transmitters, i.e., $N_1 = N_2 = N$. 
One may also use $\max \{ N_1, N_2 \}$ instead.
Algorithm~\ref{algorithm1} and Algorithm~\ref{algorithm2} solve problems (\ref{max_SR_GN}) and (\ref{max_SR_EJ}) by alternatively applying SDLC1 and SDLC2.
Therefore, we first analyze the complexity of SDLC1, which is proposed to deal with (\ref{max_SR_GN_jam_F1_v2}) and also problems in similar forms (see (\ref{max_SR_hat_F1}) - (\ref{max_SR_tilde_F2})).
The optimization of $\bm Q_1$ in SDLC1 involves matrix multiplications and eigen-decompositions, which require a complexity of ${\cal O}\left( N^3 \right)$.
In addition, the bisection search used in (\ref{optimal_A}) requires a complexity of ${\cal O}\left( N \log \left( \frac{1}{\tau} \right) \right)$, where $\tau$ is the convergence tolerance of the bisection searching method.
Therefore, the implementation of SDLC1 involves a complexity of ${\cal O}\left( N^3 + N \log \left( \frac{1}{\tau} \right) \right)$.
Analogously, we can prove that SDLC2 also requires a complexity of ${\cal O}\left( N^3 + N \log \left( \frac{1}{\tau} \right) \right)$.
By simply counting the number of times SDLC1 and SDLC2 are executed, we know that the overall complexity of Algorithm~\ref{algorithm1} and Algorithm~\ref{algorithm2} is respectively ${\cal O}\left( L_1 L_2 \left( N^3 + N \log \left( \frac{1}{\tau} \right) \right) \right)$ and ${\cal O}\left( 2 \max\{L_3, L_4\} \left( N^3 + N \log \left( \frac{1}{\tau} \right) \right) \right)$.


\section{Simulation Results}
\label{simulation}

\begin{figure}
	\centering
	\includegraphics[scale=0.47]{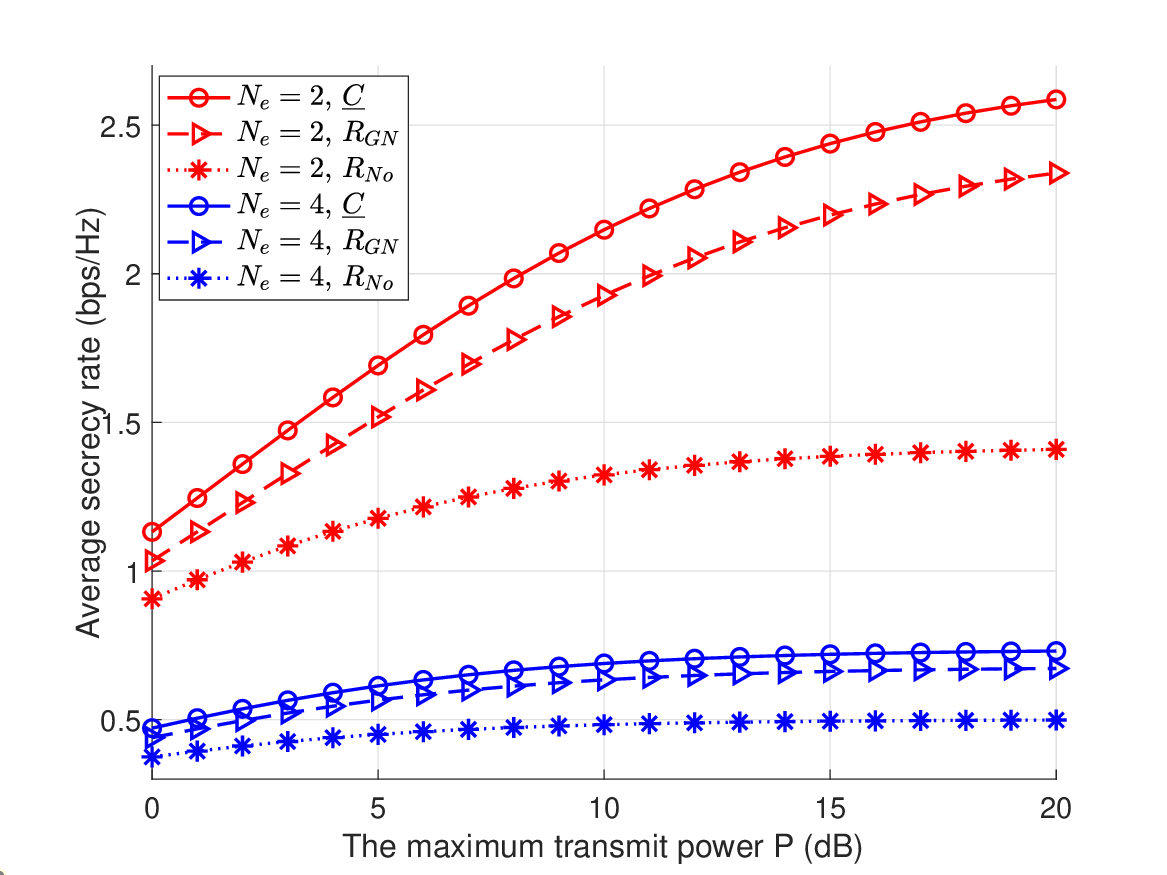}
	\caption{SIMO case: average secrecy rate obtained by different schemes versus $P$ with $N_{\text b} = 4$.}
	\label{Fig2}
\end{figure}

\begin{figure}
	\centering
	\includegraphics[scale=0.47]{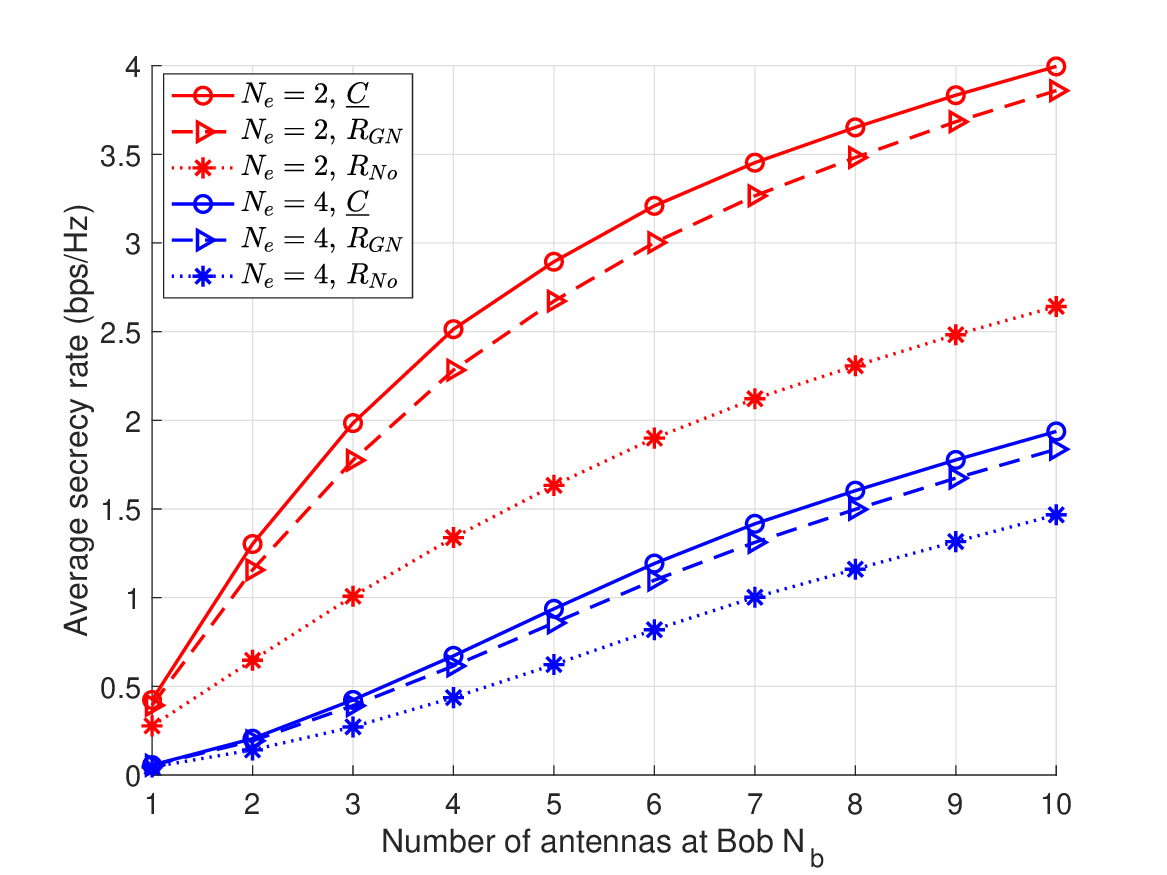}
	\caption{SIMO case: average secrecy rate obtained by different schemes versus $N_{\text b}$ with $P=20$ dB.}
	\label{Fig3}
\end{figure}

In this section, we evaluate the secrecy performance of the system by simulation.
Our main focus is on the secrecy rate ${\underline C} = \max\{R_{\text {GN}}, R_{\text {EJ}}\}$ under different parameter configurations.
For comparison, we also depict $R_{\text {GN}}$ and
\begin{equation}\label{R_No_GV}
	R_{\text {No}} \!=\! \left[ \log | \bm H_1 \bm Q_1 \bm H_1^H \!+\! \bm I_{N_{\text b}} | \!-\! \log | \bm G_1 \bm Q_1 \bm G_1^H \!+\! \bm I_{N_{\text e}} | \right]^+\!\!,\!\!
\end{equation}
which is the secrecy rate of the ``No''-jammer case, i.e., the MIMOME channel.
To maximize $R_{\text {No}}$, it can be easily verified that in the SIMO case, the optimal solution can be found, and in the MIMO case, the SDLC1 scheme can be applied.
For convenience, we assume equal maximum power constraint for both transmitters, i.e., $P_1 = P_2 = P$.
All results are obtained by averaging over $1000$ independent channel realizations.
In each realization, the entries of the channel matrices are generated according to independent and identically distributed complex Gaussian distribution with zero mean and unit variance.

\subsection{SIMO Case}

\begin{figure}
	\centering
	\includegraphics[scale=0.47]{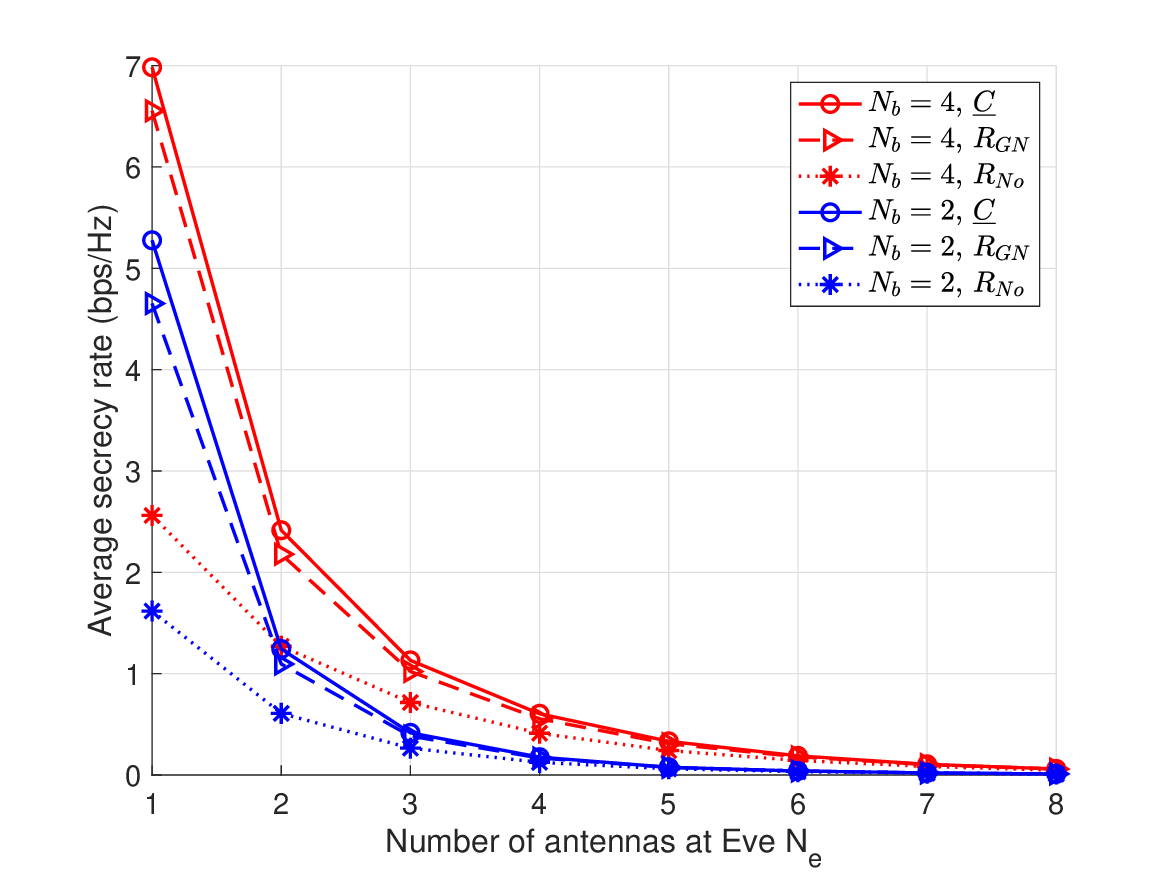}
	\caption{SIMO case: average secrecy rate obtained by different schemes versus $N_{\text e}$ with $P=20$ dB.}
	\label{Fig4}
\end{figure}

\begin{figure}
	\centering
	\includegraphics[scale=0.47]{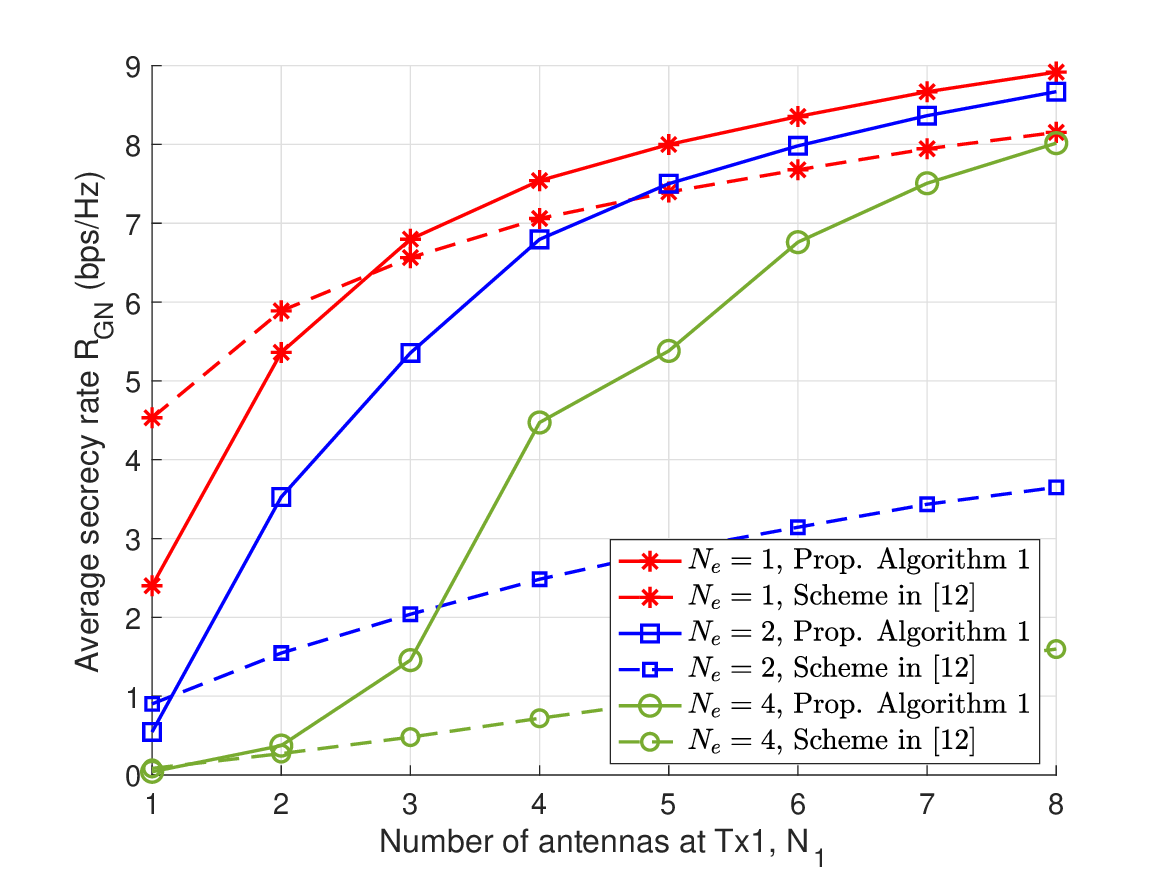}
	\caption{MISO case: average secrecy rate $R_{\text {GN}}$ obtained by different schemes versus $N_1$ with $N_{\text b} = 1$, $N_2 = 2$, and $P=20$ dB.}
	\label{Fig5}
\end{figure}

In the SIMO case, all the power control problems that aim to maximize $R_{\text {No}}$, $R_{\text {GN}}$, and ${\underline C}$, respectively, can be solved optimally.
In Figs.~\ref{Fig2}~$\sim$~\ref{Fig4}, we study the effect of different parameters $P$, $N_{\text b}$, and $N_{\text e}$ on the secrecy rate.
Several observations can be made.
First, as expected, the secrecy rate achieved by all jamming schemes increases with $N_{\text b}$ and $P$, and decreases with $N_{\text e}$.
Second, the introduction of a cooperative jammer and the utilization of either the GN or EJ scheme can lead to a significant enhancement in the secrecy performance of Tx1.
In particular, a general secrecy rate improvement of over 50\% is observed across various parameter configurations.
Furthermore, compared to the GN scheme, enabling Tx2 to switch between the GN and EJ schemes results in a noticeable increase (approximately 10\%) in the secrecy rate of Tx1.
In the following subsection, we will demonstrate that this improvement is much more substantial in the MIMO case.

\subsection{MIMO Case}

\begin{figure}
	\centering
	\includegraphics[scale=0.47]{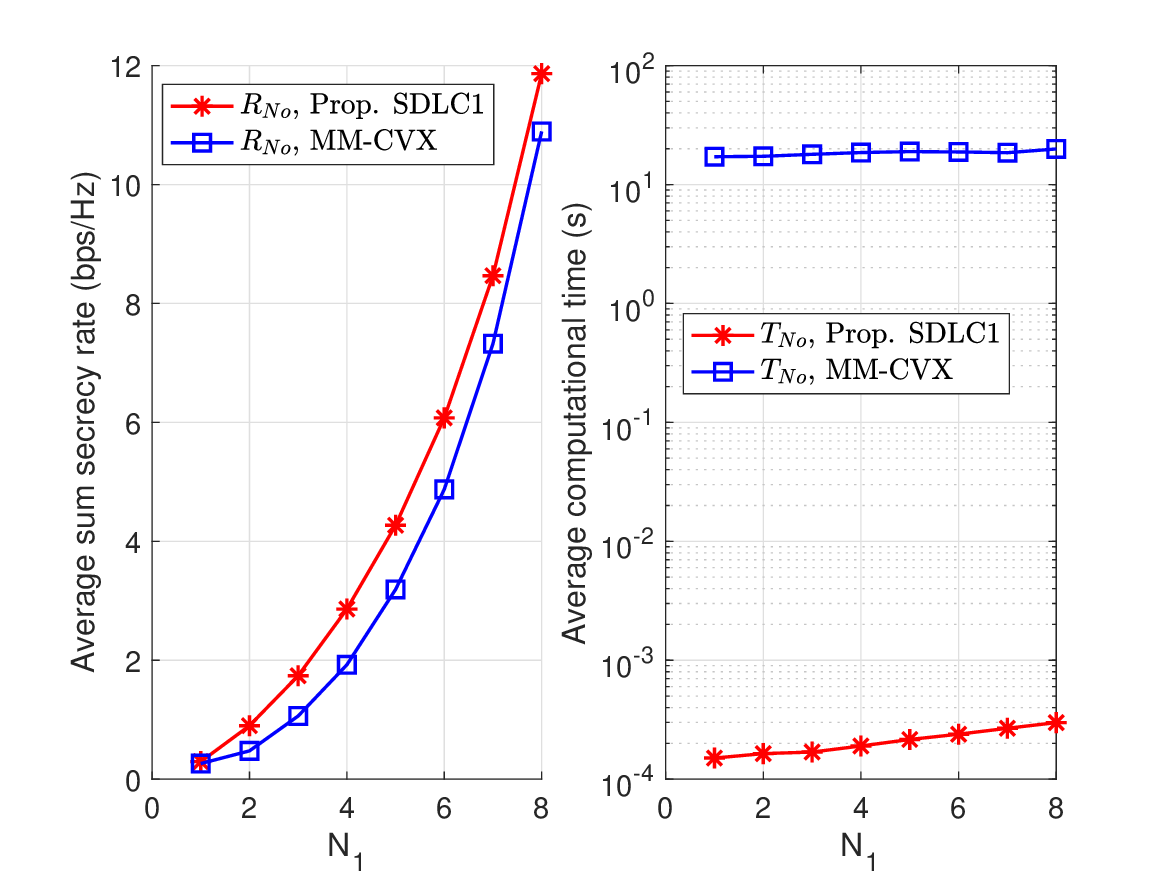}
	\caption{MIMO case with no jamming: average secrecy rate $R_{\text {No}}$ and compu-tation time $T_{\text {No}}$ versus $N_1$ with $N_{\text b} = 8$, $N_{\text e} = 8$, $N_2 = 0$, and $P=20$ dB.}
	\label{Fig6}
\end{figure}

Now we study the MIMO case.
When performing Algorithm~\ref{algorithm1} and Algorithm~\ref{algorithm2}, we set iteration numbers $L_1 = 5$, $L_2 = 50$, and $L_3 = L_4 = 5$.

In Fig.~\ref{Fig5}, we consider the MISO case, where Bob has one antenna, and compare Algorithm~\ref{algorithm1} with the scheme given in \cite{wolf2010zero}, where Tx1 and Tx2 respectively apply maximum ratio and zero-forcing transmission strategies.
It can be seen that when Eve has one antenna, i.e., $N_{\text e} = 1$, Algorithm~\ref{algorithm1} has a similar secrecy performance as the scheme provided in \cite{wolf2010zero}.
However, when $N_{\text e}$ increases (even from $1$ to $2$), Algorithm~\ref{algorithm1} significantly outperforms the scheme in \cite{wolf2010zero}, indicating that when Eve has multiple antennas, more advanced beamforming strategies should be applied to enhance security.

In Fig.~\ref{Fig6}, we consider the MIMO case with no jamming and maximize $R_{\text {No}}$ in (\ref{R_No_GV}).
Both $R_{\text {No}}$ and the computation time $T_{\text {No}}$ required for one channel realization are depicted.
We first maximize $R_{\text {No}}$ using the proposed SDLC1 scheme.
Note that as a DC programming, this problem can also be solved by the conventional MM-based technique, which first obtains a convex approximation of (\ref{R_No_GV}), and then iteratively maximizes this approximation using some standard tools, e.g., CVX, etc.
For convenience, we call this method MM-CVX.
Fig.~\ref{Fig6} shows that, in contrast to MM-CVX, the proposed SDLC1 scheme achieves better secrecy performance and, most importantly, the required computation time is several orders of magnitude lower.
In the case with a cooperative jammer, we consider problems (\ref{max_SR_GN}) and (\ref{max_SR_EJ}), and propose to solve them using Algorithm~\ref{algorithm1} and Algorithm~\ref{algorithm2}, respectively.
It can be easily verified that all subproblems (\ref{max_SR_GN_jam_F1}), (\ref{max_SR_GN_jam_F2}), and (\ref{max_SR_hat_F1}) $\sim$ (\ref{max_SR_tilde_F2}), generated in solving (\ref{max_SR_GN}) and (\ref{max_SR_EJ}), are DC programmings.
Therefore, (\ref{max_SR_GN}) and (\ref{max_SR_EJ}) can also be solved by iteratively applying MM-CVX.
However, it can be inferred from Fig.~\ref{Fig6} that this method has a similar secrecy performance in contrast to the SD-based schemes, but requires extremely high complexity since many iterations are needed.
Due to space limitation, we do not provide the comparison here.

\begin{figure}
	\centering
	\includegraphics[scale=0.47]{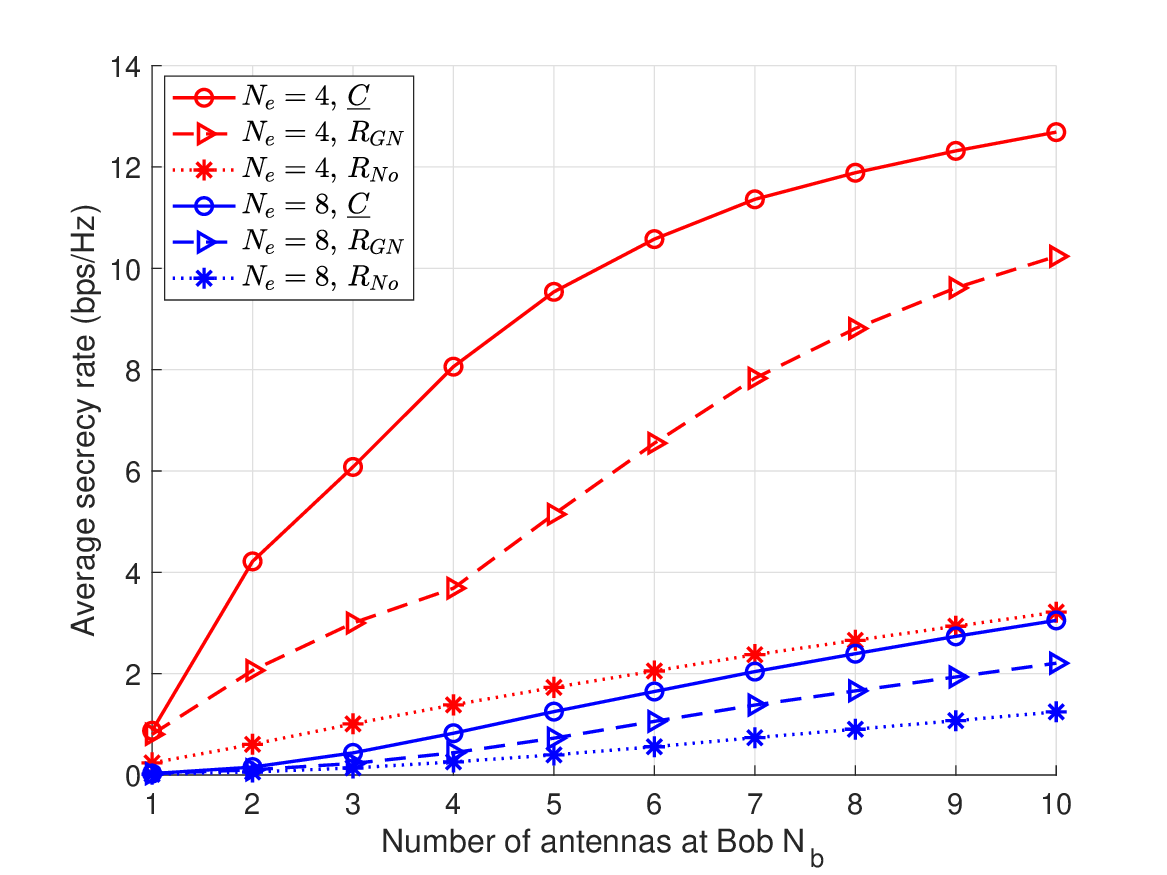}
	\caption{MIMO case: average secrecy rate obtained by different schemes versus $N_{\text b}$ with $N_1 = 2$, $N_2 = 4$, and $P = 20$ dB.}
	\label{Fig7}
\end{figure}

\begin{figure}
	\centering
	\includegraphics[scale=0.47]{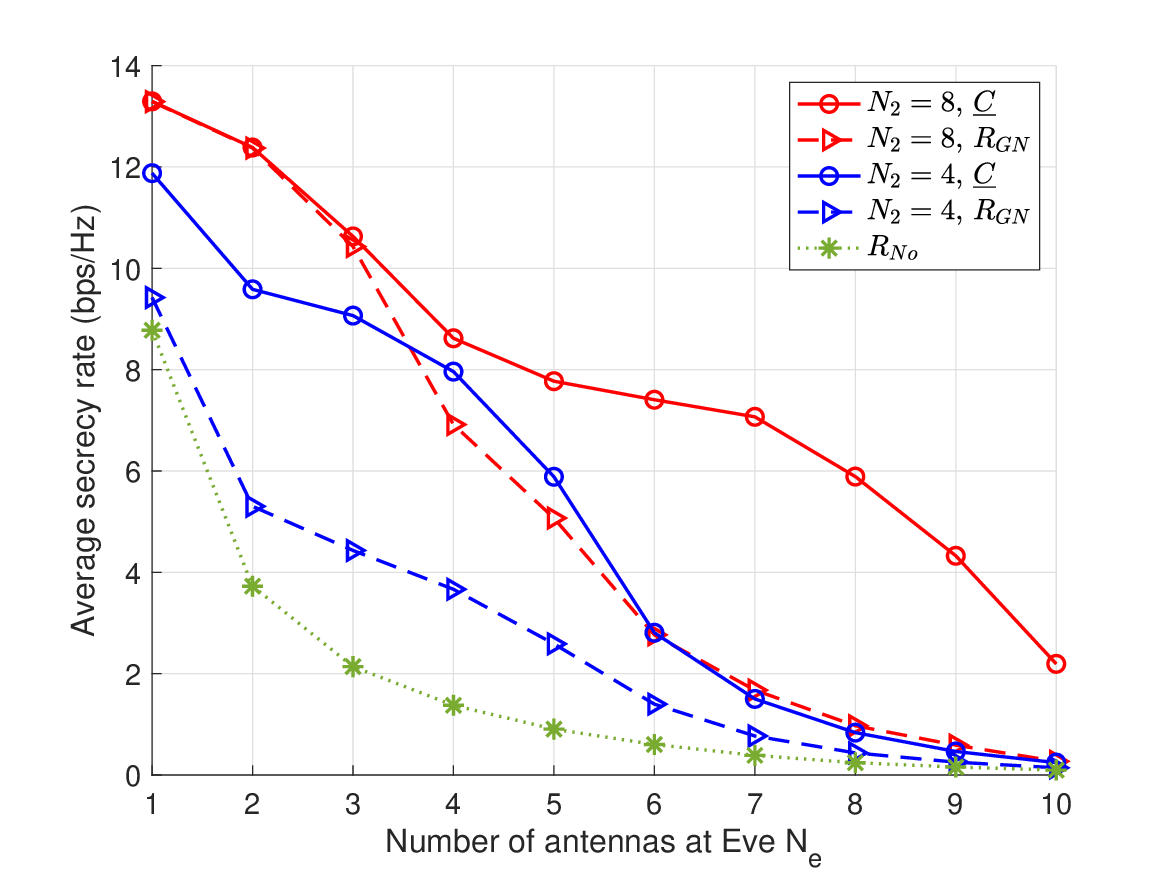}
	\caption{MIMO case: average secrecy rate obtained by different schemes versus $N_{\text e}$ with $N_{\text b} = 4$, $N_1 = 2$, and $P=20$ dB.}
	\label{Fig8}
\end{figure}

Fig.~\ref{Fig7} and Fig.~\ref{Fig8} depict the secrecy rate versus $N_{\text b}$ and $N_{\text e}$.
Comparing these two figures with Fig.~\ref{Fig3} and Fig.~\ref{Fig4}, it can be found that by deploying multiple antennas at the transmitters, the system's secrecy performance can be dramatically improved, and in addition, gaps between any two of $R_{\text {No}}$, $R_{\text {GN}}$, and ${\underline C}$ enlarge.
This observation highlights two important points.
First, jamming schemes are more effective in the MIMO case.
For example, in Fig.~\ref{Fig7}, when $N_{\text b} = N_{\text e} = 4$, we observe that $({\underline C} - R_{\text {No}}) / R_{\text {No}} > 450 \%$ and $(R_{\text {GN}} - R_{\text {No}}) / R_{\text {No}} > 150 \%$.
Second, compared to the SIMO scenario where the EJ scheme provides only a 10\% secrecy rate gain over the GN method, a substantial increase is observed in the MIMO case.
For instance, in Fig.~\ref{Fig7}, when $N_{\text b} = N_{\text e} = 4$, and in Fig.~\ref{Fig8}, when $N_2 = N_{\text e} = 4$, we both observe that $({\underline C} - R_{\text {GN}}) / R_{\text {GN}} > 100 \%$, demonstrating an excellent secrecy enhancement brought by the EJ scheme.
\begin{figure}
	\centering
	\includegraphics[scale=0.47]{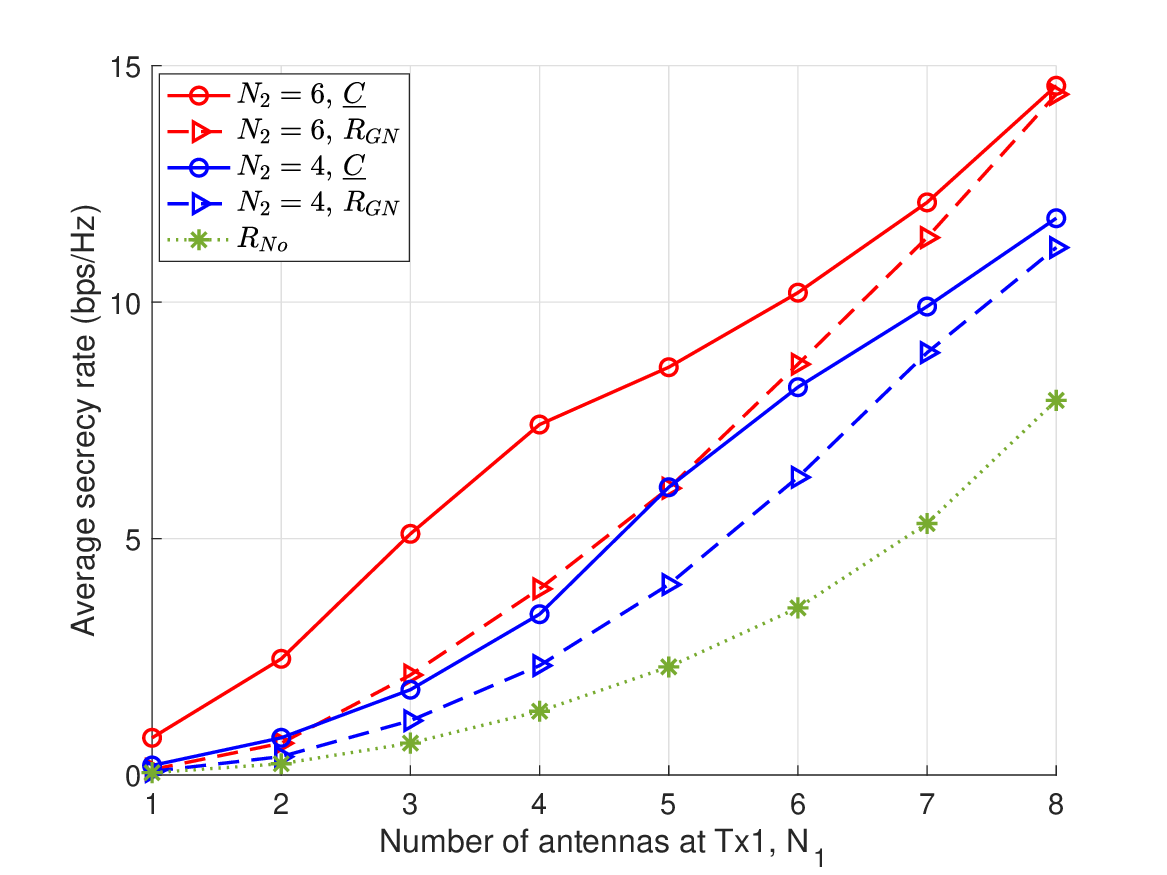}
	\caption{MIMO case: average secrecy rate obtained by different schemes versus $N_1$ with $N_{\text b} = 4$, $N_{\text e} = 8$, and $P=20$ dB.}
	\label{Fig9}
\end{figure}

\begin{figure}
	\centering
	\includegraphics[scale=0.47]{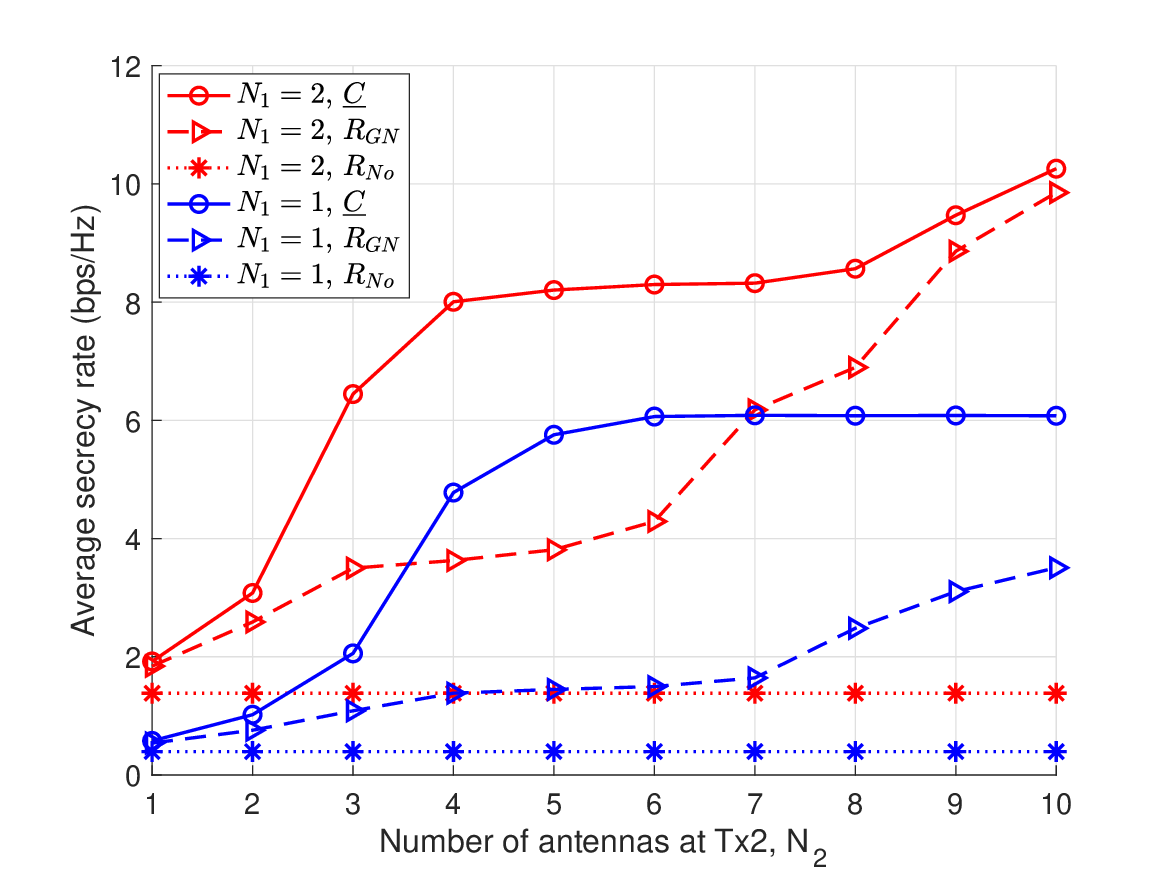}
	\caption{MIMO case: average secrecy rate obtained by different schemes versus $N_2$ with $N_{\text b} = 4$, $N_{\text e} = 4$, and $P=20$ dB.}
	\label{Fig10}
\end{figure}

It is worth noting that while ${\underline C}$ is significantly larger than $R_{\text {GN}}$ in many parameter configurations, there are still cases where ${\underline C}$ and $R_{\text {GN}}$ are relatively close. 
For example, in Fig.~\ref{Fig8}, when $N_2 = 8$ and $N_{\text e} \leq 3$, or when $N_2 = 4$ and $N_{\text e} \geq 8$, ${\underline C}$ is quite close to $R_{\text {GN}}$, indicating that in these instances, allowing Tx2 to switch between the GN and EJ schemes does not result in significant gains in the secrecy rate.
This is because if $N_{\text e}$ is small and $N_2$ is large, the interference beam of Tx2 can be designed to be narrow and precisely aligned with Eve.
Then, the jamming interference experienced by Bob is negligible or minimal.
Consequently, the additional advantage of the EJ scheme, which involves decoding the jamming signal first and subsequently canceling the interference, becomes limited.
On the contrary, if $N_{\text e}$ is large while $N_2$ is small, it indicates that Eve possesses a significant wiretapping capability, while the system's defense mechanisms may not be sufficient.
Hence, the secrecy rate is small even after introducing a cooperative jammer.
In this case, additional techniques have to be employed to enhance the secrecy, e.g., introducing more jammers, deploying more antennas at the transmitters or Bob, etc.

Fig.~\ref{Fig9} and Fig.~\ref{Fig10} investigate the effect of $N_1$ and $N_2$.
Note that for the case with no jammer, there is only Tx1.
Therefore, the value of $N_2$ does not affect that of $R_{\text {No}}$.
Similar observations regarding the improvements in secrecy achieved by introducing a cooperative jammer and employing the EJ scheme can be made from Fig.~\ref{Fig9} and Fig.~\ref{Fig10}, just as in the previous figures.
In addition, it can be seen that with fixed $N_{\text b}$ and $N_{\text e}$, the gap between ${\underline C}$ and $R_{\text {GN}}$ initially widens and then narrows as $N_1$ (or $N_2$) increases. 
This observation can be explained similarly as the phenomenon depicted in Fig.~\ref{Fig8}.
Therefore, we can conclude that by taking full advantage of the GN and EJ schemes, the secrecy performance of the system can be greatly improved, particularly in situations where the information beam cannot be precisely aligned with Bob ($N_1$ is not large enough) or the jamming beam cannot be precisely aligned with Eve ($N_2$ is not large enough).


\section{Conclusions}
\label{conclusion}

This paper studied the information-theoretic secrecy for a Gaussian MIMO wiretap channel with a cooperative jammer.
In addition to the GN jamming scheme, the jammer can also operate in the EJ mode so that its signal only interferes with Bob.
We provided an inner bound on the secrecy rate under the strong secrecy metric and aimed to maximize this bound.
We first showed that optimal power control was available for the SIMO case. 
As for the MIMO case, we developed SD-based methods with quite a low complexity to solve the problems. 
Our results showed that the secrecy performance of the system can be greatly enhanced by introducing a cooperative jammer and allowing it to switch between the GN and EJ schemes.

\appendices

\section{Proof of Theorem~\ref{theorem_inner}}
\label{prove_theorem_inner}

We prove Theorem~\ref{theorem_inner} based on the result in \cite{yassaee2010multiple} and \cite{xu2023achievable}, which respectively studied the achievable regions of two-user and $K$-user DM wiretap channels under the strong secrecy criterion.
Note that in this paper Tx1 has a secret message at rate $R$ for Bob, but Tx2 does not.
Differently, all users in \cite{yassaee2010multiple} or \cite{xu2023achievable} transmit secret messages to Bob.
To use the result, we assume that Tx2 also transmits a secret message to Bob at rate $R'$.
Then, the achievable region provided in \cite{yassaee2010multiple} for the two-user DM wiretap channel can be directly extended to the continuous Gaussian case in a standard way, by first introducing input costs and then applying discretization \cite{el2011network}.
Denote
\begin{align}\label{R1R2R3}
	& {\mathscr R}_1 (\bm Q_1, \bm Q_2) \left\{\!\!\!
	\begin{array}{ll}
		R \leq \left[ I(\bm x_1; \bm y| \bm x_2) - I(\bm x_1; \bm z) \right]^+, \\
		R' \leq \left[ I(\bm x_2; \bm y| \bm x_1) - I(\bm x_2; \bm z) \right]^+, \\
		R \!+\! R' \!\leq\! \left[ I(\bm x_1, \bm x_2; \bm y) \!-\! I(\bm x_1, \bm x_2; \bm z) \right]^+\!\!,\!\!\!\!
	\end{array} \right. 
	\\
	& {\mathscr R}_2 (\bm Q_1, \bm Q_2) \left\{
	\begin{array}{ll} 
		R' = 0, \\
		R \leq \left[ I(\bm x_1; \bm y| \bm x_2) - I(\bm x_1; \bm z| \bm x_2) \right]^+,
	\end{array} \right. 
	\\
	&{\mathscr R}_3 (\bm Q_1, \bm Q_2) \left\{
	\begin{array}{ll}
		R = 0, \\
		R' \leq \left[ I(\bm x_2; \bm y| \bm x_1) - I(\bm x_2; \bm z| \bm x_1) \right]^+,
	\end{array} \right.
\end{align}
where the mutual information expressions can be computed based on the channel model (\ref{GV_YZ}). For brevity, we only provide the detailed expressions for some of them in (\ref{R_hat_tilde_bar}).
Based on \cite[Theorem~$1$]{yassaee2010multiple} it is known that for given $\bm Q_1$ and $\bm Q_2$, any rate pair $( R, R')$ in the following region is achievable under the strong secrecy metric
\begin{equation}\label{achi_reg}
	{\mathscr R}_1 (\bm Q_1, \bm Q_2) \cup {\mathscr R}_2 (\bm Q_1, \bm Q_2) \cup {\mathscr R}_3 (\bm Q_1, \bm Q_2).
\end{equation}
Since $R$ is always $0$ in region ${\mathscr R}_3 (\bm Q_1, \bm Q_2)$, in the following, we derive (\ref{achie_region}) mainly from ${\mathscr R}_1 (\bm Q_1, \bm Q_2)$ and ${\mathscr R}_2 (\bm Q_1, \bm Q_2)$.

From the achievability proof in \cite{yassaee2010multiple} we know that to achieve a rate pair in ${\mathscr R}_1 (\bm Q_1, \bm Q_2)$, redundant messages have to be introduced to both users, i.e., besides the secret message, each user also transmits a redundant message at certain rate, such that the secret messages can be perfectly protected.
By simply setting $R' = 0$ in ${\mathscr R}_1 (\bm Q_1, \bm Q_2)$, we know that any secrecy rate $R$ satisfying 
\begin{align}\label{ub1}
	R \leq & \min \left\{ \left[ I(\bm x_1; \bm y| \bm x_2) - I(\bm x_1; \bm z) \right]^+, \right. \nonumber\\
	& \quad\quad \left. \left[ I(\bm x_1, \bm x_2; \bm y) - I(\bm x_1, \bm x_2; \bm z) \right]^+ \right\},
\end{align}
is achievable.
Note that although we set $R' = 0$, it does not imply that Tx2 transmits no message.
This can be understood by considering the $R' = \epsilon$ case, where $\epsilon$ is an arbitrarily small positive number.
Then, it is known from \cite{yassaee2010multiple} that to protect the secret message (although its rate $R'$ approaches $0$), Tx2 transmits a redundant message at certain rate such that Bob can decode the information and then eliminate the interference, but Eve cannot even if it knows the codebooks.
The EJ scheme can thus be implemented.

\setcounter{equation}{76}
\begin{figure*}[hb]
	\hrulefill
	\begin{align}\label{SR_conven_jam2_deri_p2}
		\frac{\partial R_{\text {GN}}}{\partial q_2} = \frac{a q_2^2 + b q_2 + c}{\left[ q_2 \bm h_2^H \left( P_1 \bm h_1 \bm h_1^H + \bm I_{N_{\text b}} \right)^{-1} \bm h_2 + 1 \right] \left( q_2  \bm h_2^H \bm h_2 + 1 \right) \left[ q_2 \bm g_2^H \left( P_1 \bm g_1 \bm g_1^H + \bm I_{N_{\text e}} \right)^{-1} \bm g_2 + 1 \right] \left( q_2  \bm g_2^H \bm g_2 + 1 \right) \ln 2},
	\end{align}
	\begin{align}\label{abc}
		a & = \left[ \bm h_2^H \left( P_1 \bm h_1 \bm h_1^H + \bm I_{N_{\text b}} \right)^{-1} \bm h_2 - \bm h_2^H \bm h_2 \right] \bm g_2^H \bm g_2 \bm g_2^H \left( P_1 \bm g_1 \bm g_1^H + \bm I_{N_{\text e}} \right)^{-1} \bm g_2 \nonumber\\
		& - \left[ \bm g_2^H \left( P_1 \bm g_1 \bm g_1^H + \bm I_{N_{\text e}} \right)^{-1} \bm g_2 - \bm g_2^H \bm g_2 \right] \bm h_2^H \bm h_2 \bm h_2^H \left( P_1 \bm h_1 \bm h_1^H + \bm I_{N_{\text b}} \right)^{-1} \bm h_2, \nonumber\\
		b & = 2 \left[ \bm h_2^H \left( P_1 \bm h_1 \bm h_1^H + \bm I_{N_{\text b}} \right)^{-1} \bm h_2 \bm g_2^H \bm g_2 - \bm h_2^H \bm h_2 \bm g_2^H \left( P_1 \bm g_1 \bm g_1^H + \bm I_{N_{\text e}} \right)^{-1} \bm g_2 \right], \nonumber\\
		c & = \bm h_2^H \left( P_1 \bm h_1 \bm h_1^H + \bm I_{N_{\text b}} \right)^{-1} \bm h_2 - \bm h_2^H \bm h_2 - \bm g_2^H \left( P_1 \bm g_1 \bm g_1^H + \bm I_{N_{\text e}} \right)^{-1} \bm g_2 + \bm g_2^H \bm g_2.
	\end{align}
\end{figure*}

Note that it is not always possible to guarantee that Bob can decode all the messages but Eve cannot.
For example, if the channel condition from Tx2 to Eve is much better than that from Tx2 to Bob, it is possible for Eve to decode the message of Tx2 and then eliminate the interference.
In this case, ${\mathscr R}_2 (\bm Q_1, \bm Q_2)$ implies that the following region
\setcounter{equation}{71}
\begin{equation}\label{ub2}
	R \leq  \left[ I(\bm x_1; \bm y| \bm x_2) - I(\bm x_1; \bm z| \bm x_2) \right]^+,
\end{equation}
is achievable.
Denote the upper bounds in (\ref{ub1}) and (\ref{ub2}) by
\begin{align}\label{R_hat_tilde_bar}
	{\hat R} & = \left[ I(\bm x_1; \bm y| \bm x_2) - I(\bm x_1; \bm z) \right]^+ \nonumber\\
	& = \big[ \log | \bm H_1 \bm Q_1 \bm H_1^H + \bm I_{N_{\text b}} | \nonumber\\
	& - \log | \bm G_1 \bm Q_1 \bm G_1^H \left( \bm G_2 \bm Q_2 \bm G_2^H + \bm I_{N_{\text e}} \right)^{-1} + \bm I_{N_{\text e}} | \big]^+,\nonumber\\
	{\tilde R} & = \left[ I(\bm x_1, \bm x_2; \bm y) - I(\bm x_1, \bm x_2; \bm z) \right]^+ \nonumber\\
	& = \left[ \log | \bm H_1 \bm Q_1 \bm H_1^H + \bm H_2 \bm Q_2 \bm H_2^H + \bm I_{N_{\text b}} | \right.\nonumber\\
	& \left. - \log | \bm G_1 \bm Q_1 \bm G_1^H + \bm G_2 \bm Q_2 \bm G_2^H + \bm I_{N_{\text e}} | \right]^+, \nonumber\\
	{\bar R} & = \left[ I(\bm x_1; \bm y| \bm x_2) - I(\bm x_1; \bm z| \bm x_2) \right]^+ \nonumber\\
	& = \left[ \log | \bm H_1 \bm Q_1 \bm H_1^H \!+\! \bm I_{N_{\text b}} | \!-\! \log | \bm G_1 \bm Q_1 \bm G_1^H \!+\! \bm I_{N_{\text e}} | \right]^+\!\!.\!\!
\end{align}
Then, any secrecy rate satisfying $R \!\leq\! R_{\text {EJ}} \triangleq \max \{ \min \{ {\hat R}, {\tilde R} \},$ ${\bar R} \}$ is achievable.
Theorem~\ref{theorem_inner} is thus proven.

\section{Proof of Theorem~\ref{theorem_opt_solu_conven}}
\label{prove_theorem_opt_solu_conven}

Using (\ref{O1O2_1}), $R_{\text {GN}}$ in (\ref{SR_conven_jam}) can be rewritten as
\begin{align}\label{SR_conven_jam2}
R_{\text {GN}} & = \log \left[ q_1 \bm h_1^H \left( q_2 \bm h_2 \bm h_2^H + \bm I_{N_{\text b}} \right)^{-1} \bm h_1 + 1 \right] \nonumber\\
& - \log \left[ q_1 \bm g_1^H \left( q_2 \bm g_2 \bm g_2^H + \bm I_{N_{\text e}} \right)^{-1} \bm g_1 + 1 \right],
\end{align}
where $[\cdot]^+$ is omitted for convenience.
Its first-order partial derivation over $q_1$ is
\begin{align}\label{SR_conven_jam2_deri_p1}
\frac{\partial R_{\text {GN}}}{\partial q_1} = & \frac{1}{\ln 2} \left[  \frac{\bm h_1^H \left( q_2 \bm h_2 \bm h_2^H + \bm I_{N_{\text b}} \right)^{-1} \bm h_1}{q_1 \bm h_1^H \left( q_2 \bm h_2 \bm h_2^H + \bm I_{N_{\text b}} \right)^{-1} \bm h_1 + 1 } \right. \nonumber\\
& \quad\quad \left. - \frac{\bm g_1^H \left( q_2 \bm g_2 \bm g_2^H + \bm I_{N_{\text e}} \right)^{-1} \bm g_1}{ q_1 \bm g_1^H \left( q_2 \bm g_2 \bm g_2^H + \bm I_{N_{\text e}} \right)^{-1} \bm g_1 + 1} \right].
\end{align}
It can be easily checked that whether (\ref{SR_conven_jam2_deri_p1}) is positive or not is determined by the values of $\bm h_1^H \left( q_2 \bm h_2 \bm h_2^H + \bm I_{N_{\text b}} \right)^{-1} \bm h_1$ and $\bm g_1^H \left( q_2 \bm g_2 \bm g_2^H + \bm I_{N_{\text e}} \right)^{-1} \bm g_1$.
Hence, for a given $q_2$, $R_{\text {GN}}$ either increases or decreases with $q_1$ in $[0, P_1]$.
The optimal $q_1$ is thus either $0$ or $P_1$.
Since $R_{\text {GN}} = 0$ if $q_1 = 0$, we only need to talk about the case with $q_1 = P_1$.
If $q_1 = P_1$, using (\ref{O1O2_1}) and (\ref{O1O2_2}), $R_{\text {GN}}$ in (\ref{SR_conven_jam}) can be rewritten as (without $[\cdot]^+$)
\begin{align}\label{SR_conven_jam3}
R_{\text {GN}} & = \log | q_2 \bm h_2 \bm h_2^H \left( P_1 \bm h_1 \bm h_1^H + \bm I_{N_{\text b}} \right)^{-1} + \bm I_{N_{\text b}} | \nonumber\\
& + \log | P_1 \bm h_1 \bm h_1^H + \bm I_{N_{\text b}} | - \log | q_2 \bm h_2  \bm h_2^H + \bm I_{N_{\text b}} | \nonumber\\
& - \log | q_2 \bm g_2 \bm g_2^H \left( P_1 \bm g_1 \bm g_1^H + \bm I_{N_{\text e}} \right)^{-1} + \bm I_{N_{\text e}} | \nonumber\\
& - \log | P_1 \bm g_1 \bm g_1^H + \bm I_{N_{\text e}} | + \log | q_2 \bm g_2 \bm g_2^H + \bm I_{N_{\text e}} | \nonumber\\
& = \log \left[ q_2 \bm h_2^H \left( P_1 \bm h_1 \bm h_1^H + \bm I_{N_{\text b}} \right)^{-1} \bm h_2 + 1 \right] \nonumber\\
& + \log \left( P_1 \bm h_1^H \bm h_1 + 1 \right) - \log \left( q_2  \bm h_2^H \bm h_2 + 1 \right) \nonumber\\
& - \log \left[ q_2 \bm g_2^H \left( P_1 \bm g_1 \bm g_1^H + \bm I_{N_{\text e}} \right)^{-1} \bm g_2 + 1 \right] \nonumber\\
& - \log \left( P_1 \bm g_1^H \bm g_1 + 1 \right) + \log \left( q_2  \bm g_2^H \bm g_2 + 1 \right).
\end{align}
Its first-order partial derivation over $q_2$, i.e., $\partial R_{\text {GN}} / \partial q_2$ is given in (\ref{SR_conven_jam2_deri_p2}) at the bottom of this page, in which $a$, $b$, and $c$ are defined in (\ref{abc}).
The optimal $q_2$ can then be easily found by talking about the values of $a$, $b$, and $c$.

If $a = 0$, $b < 0$, and $0 < - \frac{c}{b} < P_2$, the zero point of (\ref{SR_conven_jam2_deri_p2}) is $- \frac{c}{b}$ and it can be easily checked that $R_{\text {GN}}$ increases with $q_2$ in $\left[ 0, - \frac{c}{b} \right]$ and decreases in $\left[ - \frac{c}{b}, P_2 \right]$.
Hence, $q_2^* = - \frac{c}{b}$.

If $a > 0$ and $b^2 - 4ac > 0$, the parabola $a q_2^2 + b q_2 + c$ opens upward and has zero points
\setcounter{equation}{78}
\begin{align}\label{zero_point}
P_0 & = \frac{-b - \sqrt{b^2 - 4ac}}{2 a}, \nonumber\\
{\hat P}_0 & = \frac{-b + \sqrt{b^2 - 4ac}}{2 a},
\end{align}
with $P_0 < {\hat P}_0$.
If $P_0 > 0$, $R_{\text {GN}}$ increases with $q_2$ in $\left[ 0, P_0 \right]$ and $\left[ {\hat P}_0, + \infty \right]$ and decreases in $\left[ P_0, {\hat P}_0 \right]$.
Hence, $R_{\text {GN}} (P_1, P_0) > R_{\text {GN}} (P_1, q_2), \forall q_2 \in [ 0, P_0 )$.
Since $q_2 \leq P_2$, it can be easily verified that if $0 < P_0 < P_2$, the optimal $q_2^*$ that maximizes $R_{\text {GN}} (P_1, q_2)$ is either $P_0$ or $P_2$.

If $a < 0$ and $b^2 - 4ac > 0$, the parabola $a q_2^2 + b q_2 + c$ opens downward and also has the two zero points in (\ref{zero_point}), but with ${\hat P}_0 < P_0$.
In this case, if $0 < P_0 < P_2$, $R_{\text {GN}}$ decreases with $q_2$ in $\left[ P_0, P_2 \right]$.
Hence, $R_{\text {GN}} (P_1, P_0) > R_{\text {GN}} (P_1, q_2), \forall q_2 \in ( P_0, P_2 ]$.
The monotonicity of $R_{\text {GN}}$ w.r.t. $q_2$ in $[0, P_0]$ depends on whether ${\hat P}_0$ is in $[0, P_0]$ or not.
If ${\hat P}_0 \leq 0$, $R_{\text {GN}}$ increases monotonically with $q_2$ in $[0, P_0]$.
Otherwise, it decreases first in $[0, {\hat P}_0]$ and then increases in $[{\hat P}_0, P_0]$.
Hence, the optimal $q_2^*$ that maximizes $R_{\text {GN}} (P_1, q_2)$ is either $0$ or $P_0$.

For all the other cases, it can be easily checked that the optimal $q_2^*$ that maximizes $R_{\text {GN}} (P_1, q_2)$ is either $0$ or $P_2$.
We omit the details here due to space limitation.

\section{Proof of Theorem~\ref{theorem_opt_p1p2_new}}
\label{prove_theorem_opt_p1p2_new}

We first solve problem (\ref{sub_problem1}).
Using (\ref{O1O2_1}), ${\hat R}$ in (\ref{SR_hat_tilde_bar_SIMO}) can be rewritten as
\begin{equation}\label{SR_hat2}
{\hat R} \!=\! \log \!\left( q_1 \bm h_1^H \bm h_1 \!+\! 1 \right) \!-\! \log \left[ q_1 \bm g_1^H \!\!\left( q_2 \bm g_2 \bm g_2^H \!+\! \bm I_{N_{\text e}} \right)^{\!-1} \!\!\bm g_1 \!+\! 1 \right]\!\!,
\end{equation}
where $[\cdot]^+$ is omitted for convenience.
It can be easily proven by using eigen-decomposition that $\bm g_1^H \left( q_2 \bm g_2 \bm g_2^H + \bm I_{N_{\text e}} \right)^{-1} \bm g_1$ decreases with $q_2$.
Hence, ${\hat R}$ increases with $q_2$ and in the optimal case of (\ref{sub_problem1}), ${\hat q}_2^* = P_2$.
With ${\hat q}_2^* = P_2$, the first-order partial derivation over $q_1$ is
\begin{equation}\label{SR_hat2_deri_p1}
\frac{{\hat R}}{\partial q_1} \!=\! \frac{1}{\ln 2} \!\left[\!  \frac{\bm h_1^H \bm h_1}{ q_1 \bm h_1^H \bm h_1 \!+\! 1 } \!-\! \frac{ \bm g_1^H \left( P_2 \bm g_2 \bm g_2^H + \bm I_{N_{\text e}} \right)^{-1} \bm g_1 }{ q_1 \bm g_1^H \!\left( P_2 \bm g_2 \bm g_2^H \!+\! \bm I_{N_{\text e}} \right)^{\!-1} \!\!\bm g_1 \!+\! 1 } \!\right]\!\!.
\end{equation}
Similar to (\ref{SR_conven_jam2_deri_p1}), (\ref{SR_hat2_deri_p1}) shows that the optimal $q_1$ should be either ${\hat P}_{1, {\text {lb}}}$ or ${\hat P}_{1, {\text {ub}}}$, depending on the values of $\bm h_1^H \bm h_1$ and $\bm g_1^H \left( P_2 \bm g_2 \bm g_2^H + \bm I_{N_{\text e}} \right)^{-1} \bm g_1$.
We give the solution in (\ref{opt_p1_new_jam}).

Next, we solve problem (\ref{sub_problem2}).
Using (\ref{O1O2_1}) and (\ref{O1O2_2}), ${\tilde R}$ in (\ref{SR_hat_tilde_bar_SIMO}) can be equivalently rewritten as
\begin{align}\label{SR_tilde2}
& {\tilde R} =\! \log \!\left[ q_1 \bm h_1^H \!\!\left( q_2 \bm h_2 \bm h_2^H \!\!+\! \bm I_{N_{\text b}} \right)^{\!-1} \!\bm h_1 \!+\! 1 \right] \!+\! \log\! \left( q_2 \bm h_2^H \bm h_2 \!+\! 1 \right) \nonumber\\ 
& -\! \log\! \left[\! q_1 \bm g_1^H \!\!\left( q_2 \bm g_2 \bm g_2^H \!\!+\! \bm I_{N_{\text e}} \right)^{\!-1} \!\!\bm g_1 \!+\! 1 \!\right] \!-\! \log \!\left( q_2 \bm g_2^H \bm g_2 \!+\! 1 \right)\!.\!\!\!
\end{align}
Its first-order partial derivation over $q_1$ is
\begin{align}\label{SR_tilde2_deri_p1}
\frac{{\tilde R}}{\partial q_1} = & \frac{1}{\ln 2} \left[  \frac{\bm h_1^H \left( q_2 \bm h_2 \bm h_2^H + \bm I_{N_{\text b}} \right)^{-1} \bm h_1}{q_1 \bm h_1^H \left( q_2 \bm h_2 \bm h_2^H + \bm I_{N_{\text b}} \right)^{-1} \bm h_1 + 1 } \right. \nonumber\\ 
&\quad\quad - \left. \frac{\bm g_1^H \left( q_2 \bm g_2 \bm g_2^H + \bm I_{N_{\text e}} \right)^{-1} \bm g_1}{ q_1 \bm g_1^H \left( q_2 \bm g_2 \bm g_2^H + \bm I_{N_{\text e}} \right)^{-1} \bm g_1 + 1} \right],
\end{align}
indicating that in the optimal case of (\ref{sub_problem2}), $q_1$ should be either ${\tilde P}_{1, {\text {lb}}}$ or ${\tilde P}_{1, {\text {ub}}}$.
Similarly, we can prove that the optimal $q_2$ should be either $0$ or $P_2$.
Therefore, as shown in (\ref{opt_p2_new_jam}), the optimal solution of (\ref{sub_problem2}) can be easily found by checking four possible solutions.

\section{Proof of Theorem~\ref{theorem_opt_solu_new}}
\label{prove_theorem_opt_solu_new}

Based on (\ref{SR_hat_tilde_bar_SIMO}), the difference between ${\hat R}$ and ${\tilde R}$ (without $[\cdot]^+$) is
\begin{align}\label{SR_hat_SR_tilde_diff}
& {\hat R} - {\tilde R} \nonumber\\
= & \log | q_1 \bm h_1 \bm h_1^H + \bm I_{N_{\text b}} | + \log | q_2 \bm g_2 \bm g_2^H + \bm I_{N_{\text e}} | \nonumber\\
& - \log | q_1 \bm h_1 \bm h_1^H + q_2 \bm h_2 \bm h_2^H + \bm I_{N_{\text b}} | \nonumber\\
= & \log\! \left( q_2 \bm g_2^H \bm g_2 \!\!+\!\! 1 \right) \!-\! \log\! \left[\! q_2 \bm h_2^H \!\left( q_1 \bm h_1 \bm h_1^H \!\!+\! \bm I_{N_{\text b}} \right)^{\!\!-\!1} \!\bm h_2 \!+\! 1 \!\right]\!,\!\!\!\!
\end{align}
where the equations holds due to (\ref{O1O2_1}) and (\ref{O1O2_2}).
It is obvious from (\ref{SR_hat_SR_tilde_diff}) that the sign of ${\hat R} - {\tilde R}$ is determined by the relative magnitudes of $\bm g_2^H \bm g_2$ and $\bm h_2^H \left( q_1 \bm h_1 \bm h_1^H + \bm I_{N_{\text b}} \right)^{-1} \bm h_2$.
Since $\bm h_2^H \left( q_1 \bm h_1 \bm h_1^H + \bm I_{N_{\text b}} \right)^{-1} \bm h_2$ decreases with $q_1$, we have 
\begin{equation}
\bm h_2^H \left( P_1 \bm h_1 \bm h_1^H + \bm I_{N_{\text b}} \right)^{-1} \bm h_2 \leq \bm h_2^H \bm h_2.
\end{equation}

In the following we prove Theorem~\ref{theorem_opt_solu_new} by separately discussing three possible cases.
First, if $\bm g_2^H \bm g_2 \leq \bm h_2^H \left( P_1 \bm h_1 \bm h_1^H + \bm I_{N_{\text b}} \right)^{-1} \bm h_2$, it is known from (\ref{SR_hat_SR_tilde_diff}) that
\begin{equation}\label{case1}
{\hat R} \leq {\tilde R}, ~\forall~ q_1 \in [0, P_1].
\end{equation}
Problem (\ref{max_SR_new_jam2}) can thus be optimally solved by letting ${\hat P}_{1, {\text {lb}}} = 0$ and ${\hat P}_{1, {\text {ub}}} = P_1$, and dealing with (\ref{sub_problem1}).

Second, if $\bm g_2^H \bm g_2 \geq \bm h_2^H \bm h_2$, we have
\begin{equation}\label{case2}
{\hat R} \geq {\tilde R}, ~\forall~ q_1 \in [0, P_1].
\end{equation}
Problem (\ref{max_SR_new_jam2}) can thus be optimally solved by letting ${\tilde P}_{1, {\text {lb}}} = 0$ and ${\tilde P}_{1, {\text {ub}}} = P_1$, and dealing with (\ref{sub_problem2}).

Third, if $\bm h_2^H \left( P_1 \bm h_1 \bm h_1^H + \bm I_{N_{\text b}} \right)^{-1} \bm h_2 < \bm g_2^H \bm g_2 < \bm h_2^H \bm h_2$, since $\bm h_2^H \left( q_1 \bm h_1 \bm h_1^H + \bm I_{N_{\text b}} \right)^{-1} \bm h_2$ monotonically decreases with $q_1$, there must exist a point $P_0' \in (0, P_1)$ such that
\begin{equation}\label{eq1}
\bm h_2^H \left( P_0' \bm h_1 \bm h_1^H + \bm I_{N_{\text b}} \right)^{-1} \bm h_2 = \bm g_2^H \bm g_2.
\end{equation}
It is obvious that the matrix $P_0' \bm h_1 \bm h_1^H + \bm I_{N_{\text b}}$ is positive definite and its eigenvalues are all one except $P_0' \bm h_1^H \bm h_1 + 1$.
Denote its eigen-decomposition by
\begin{equation}\label{eigen_decomp1}
P_0' \bm h_1 \bm h_1^H + \bm I_{N_{\text b}} = \bm L {\text {diag}} \{1, \cdots, 1, P_0' \bm h_1^H \bm h_1 + 1\} \bm L^H,
\end{equation}
and let
\begin{equation}\label{h2_hat}
{\hat {\bm h}}_2 = \bm L^H \bm h_2 \triangleq [ {\hat h}_{2, 1}, \cdots, {\hat h}_{2, N_{\text b}} ]^T.
\end{equation}
Based on (\ref{eigen_decomp1}) and (\ref{h2_hat}), (\ref{eq1}) can be rewritten as
\begin{align}
& {\hat {\bm h}}_2^H {\text {diag}} \left\{1, \cdots, 1, \frac{1}{P_0' \bm h_1^H \bm h_1 + 1} \right\} {\hat {\bm h}}_2 \nonumber\\
= & \sum_{i = 1}^{N_{\text b} - 1} |{\hat h}_{2, i}|^2 + \frac{|{\hat h}_{2, N_{\text b}}|^2}{P_0' \bm h_1^H \bm h_1 + 1} \nonumber\\
= & \bm g_2^H \bm g_2,
\end{align}
from which we get
\begin{equation}\label{P_0}
P_0' = \frac{|{\hat h}_{2, N_{\text b}}|^2}{ \bm h_1^H \bm h_1 \left( \bm g_2^H \bm g_2 - \sum_{i = 1}^{N_{\text b} - 1} |{\hat h}_{2, i}|^2 \right) } - \frac{1}{\bm h_1^H \bm h_1}.
\end{equation}
Then, we know that
\begin{align}\label{case3}
{\hat R} & \leq {\tilde R}, ~\forall~ q_1 \in [0, P_0'], \nonumber\\
{\hat R} & \geq {\tilde R}, ~\forall~ q_1 \in [P_0', P_1].
\end{align}
According to (\ref{case3}), problem (\ref{max_SR_new_jam2}) can be solved by considering two cases.
If $q_1 \in [0, P_0']$, we let ${\hat P}_{1, {\text {lb}}} = 0$ and ${\hat P}_{1, {\text {ub}}} = P_0'$, and deal with (\ref{sub_problem1}).
If $q_1 \in [P_0', P_1]$, we let ${\tilde P}_{1, {\text {lb}}} = P_0'$ and ${\tilde P}_{1, {\text {ub}}} = P_1$, and deal with (\ref{sub_problem2}).
The optimal solution of (\ref{max_SR_new_jam2}) is thus (\ref{solu_case3}).
This completes the proof.

\section{Proof of Theorem~\ref{theo_opt_lambda}}
\label{prove_theo_opt_lambda}

Since problem (\ref{SD_problem3}) is convex and has affine constraints, the strong duality holds and its optimal solution can be obtained by checking the KKT condition.
Attaching a Lagrange multiplier $\beta$ to the constraint (\ref{SD_problem3_c}), we get the following Lagrange function
\begin{align}\label{Lagrange_2}
	{\cal L} (\bm \varLambda, \beta) & = \sum_{n=1}^{{\hat N}_2} \left[ - \ln (\rho_n \lambda_n + 1) - \ln ((1 - \rho_n) \lambda_n + 1) \right] \nonumber\\
	& + \sum_{n=1}^{N_2} (a_n + \beta \left\| \bm w_n \right\|^2 ) \lambda_n - \beta P_2.
\end{align}
When ${\hat N}_2 + 1 \leq n \leq N_2$, the objective function (\ref{SD_problem3_a}) is non-decreasing w.r.t. $\lambda_n$.
Hence, 
\begin{equation}\label{lambda_case1}
	\lambda_n^* = 0, ~{\text {if}}~ {\hat N}_2 + 1 \leq n \leq N_2.
\end{equation}
If $1 \leq n \leq {\hat N}_2 + 1$ and $\rho_n = 0$ or $\rho_n = 1$, the first-order partial derivation of the Lagrange function ${\cal L}$ over $\lambda_n$ is
\begin{equation}\label{deri_case2}
	\frac{\partial {\cal L}}{\partial \lambda_n} = - \frac{1}{\lambda_n + 1} + a_n + \beta \left\| \bm w_n \right\|^2.
\end{equation}
By checking the first-order optimality condition and ensuring $\lambda_n \geq 0$, we have
\begin{equation}\label{lambda_case2}
	\lambda_n^* (\beta^*) \!=\! \left[ \frac{1}{a_n \!+\! \beta^* \left\| \bm w_n \right\|^2} \!-\! 1 \right]^{\!+}\!\!\!, ~{\text {if}}~ 1 \!\leq\! n \!\leq\! {\hat N}_2 ~{\text {and}}~ \rho_n \!=\! 0 ~{\text {or}}~ 1.
\end{equation}
If $1 \leq n \leq {\hat N}_2 + 1$ and $0 < \rho_n < 1$, the first-order partial derivation of ${\cal L}$ over $\lambda_n$ is
\begin{align}\label{deri_case3}
	& \frac{\partial {\cal L}}{\partial \lambda_n} = \nonumber\\
	& \frac{(a_n \!+\! \beta \left\| \bm w_n \right\|^2) [ \rho_n (1 \!-\! \rho_n) \lambda_n^2 \!+\! \lambda_n \!+\! 1 ] \!-\! 2 \rho_n (1 \!-\! \rho_n) \lambda_n \!-\! 1}{(\rho_n \lambda_n + 1) [(1 - \rho_n) \lambda_n + 1] }.
\end{align}
By checking the first-order optimality condition and ensuring $\lambda_n \geq 0$, we have
\begin{align}\label{lambda_case3}
	\lambda_n^* (\beta^*) \!= & \frac{\left[\! - 1 \!+\! \frac{2 \rho_n (1 - \rho_n)}{a_n \!+\! \beta^* \left\| \bm w_n \right\|^2} \!+\! \sqrt{\! (2 \rho_n \!\!-\!\! 1)^2 \!+\! \frac{4 \rho_n^2 (1 \!-\! \rho_n)^2}{\left( a_n + \beta^* \left\| \bm w_n \right\|^2 \right)^2}} \right]^+}{2 \rho_n (1 - \rho_n)}, \nonumber\\
	& {\text {if}}~ 1 \leq n \leq {\hat N}_2 ~{\text {and}}~ 0 < \rho_n < 1.
\end{align}
It is known from the KKT condition that in the optimal case of (\ref{SD_problem3}),
\begin{equation}\label{KKT}
	\beta^* \left( \sum_{n=1}^{N_2} \left\| \bm w_n \right\|^2 \lambda_n^* - P_2 \right) = 0.
\end{equation}
Therefore, if $\lambda_n^* (0)$ makes the constraint (\ref{SD_problem3_c}) hold, we have $\beta^* = 0$.
Otherwise, $\beta^* > 0$ and $\beta^*$ can be found using the bisection searching method such that (\ref{SD_problem3_c}) holds with equality since $\lambda_n (\beta)$ decreases with $\beta$.
This completes the proof.

\bibliographystyle{IEEEtran}
\bibliography{IEEEabrv,Ref}

\end{document}